\documentclass[aps,nopacs,showkeys,nofootinbib,preprint]{revtex4}
\usepackage{epsfig,amssymb,bm,graphicx,color,amsmath,rotating}
\usepackage{xcolor}
\usepackage[percent]{overpic}
\usepackage{mwe,xcolor}
\usepackage{transparent}
\usepackage{amsmath}
\usepackage{framed}
\usepackage{slashed}
\usepackage[export]{adjustbox}
\usepackage{enumerate} 
\usepackage{hyperref}

\newcommand{\comment}[1]{}

\begin{document} {\normalsize}

\interfootnotelinepenalty=10000

\title{Strong-field QED in Furry-picture momentum-space formulation: \\
Ward identities and Feynman diagrams} 

\author{U. Hernandez Acosta${}^{1, 2}$, B.~K\"ampfer${}^{1, 3}$}
\affiliation{${}^{1}$Helmholtz-Zentrum  Dresden-Rossendorf, 01328 Dresden, Germany}
\affiliation{${}^{2}$Center for Advanced Systems Understanding (CASUS), Helmholtz-Zentrum
Dresden-Rossendorf, Untermarkt 20, 02826 G\"orlitz, Germany}
\affiliation{${}^{3}$Institut f\"ur Theoretische Physik, TU~Dresden, 01062 Dresden, Germany}
 
\begin{abstract}
The impact of a strong electromagnetic background field on otherwise perturbative QED processes
is studied in the momentum-space formulation. The univariate background field is assumed 
to have finite support in time, thus being suitable to provide a model for a strong laser pulse
in plane-wave approximation. The usually employed Furry picture in position space
must be equipped with some non-obvious terms to ensure the Ward identity. In contrast,
the momentum space formulation allows for an easy and systematic account of these terms,
both globally and order-by-order in the weak-field expansion.
In the limit of an infinitely long-acting (monochromatic) background field, these terms become gradually suppressed,
and the standard perturbative QED Feynman diagrams are recovered in the leading-order weak-field limit. 
A few examples of three- and four-point amplitudes are considered to demonstrate the application of our
Feynman rules which employ free Dirac spinors, the free photon propagator, and the free Fermion propagator,
while the external field impact is solely encoded in the Fermion-Fermion-photon vertex function.  
The appearance of on-/off-shell contributions, singular structures, and Oleinik resonances is pointed out.
  
\end{abstract}


\date{\today}

\maketitle



\section{Introduction} \label{introduction}

With the advent of permanently increasing laser intensities by advanced technologies
\cite{world_record} one gets access towards the strong-field regime of QED \cite{Marklund:2022gki}
in a hitherto uncharted region
of parameter space. Among the famous examples are the ``vacuum break-down" by
the Sauter-Schwinger effect (cf.\ \cite{Ilderton:2021zej,Kohlfurst:2021skr,Sevostyanov:2020dhs,Taya:2020dco} 
for recent activities and entries to extensive citations)
and the unsettled implications \cite{Heinzl:2021mji,Edwards:2020npu,Ilderton:2019kqp} of the Ritus-Narozhny conjecture 
(cf.\ \cite{Fedotov:2016afw} for an introduction).
While QED delivers unprecedentedly accurate results in certain regions of the parameter space,
at high energies of the involved particles and for processes in strong background fields there is still
room for testing the theory for ``physics beyond the Standard Model" or verifying long-standing
predictions. For instance, the soft-photon theorems - at the heart of the IR structure of QED -
seem to fail when many hadrons are involved in the final state of a high-energy strong-interaction
collisions. This issue triggered activities for new detector concepts and plans of experimental
investigations at the LHC \cite{Adamova:2019vkf}. 
In fact, Strominger's IR triangle \cite{Strominger:2017zoo} 
finds recently much interest
culminating in ``new symmetries of QED" \cite{Kapec:2015ena}, see e.g. \cite{Feal:2022iyn}.

The current standard approach to calculations of QED processes in strong background fields deploys
the Furry picture in position space.
Fermions (electrons and positrons) are dressed by accounting
exactly for the (quasi-classical) interaction with the electromagnetic background field, while the interaction 
of Fermions with photons is dealt with in order-$\alpha$ perturbative expansion, see 
\cite{Fedotov:2022ely,Gonoskov:2021hwf,DiPiazza:2011tq} for reviews.
A convenient model of the laser field is accomplished in the plane-wave approximation. 
The external field, taken as given background (for back-reaction dynamics, cf.\ \cite{Ilderton:2017xbj,Seipt:2016fyu}),
depends then only on one variable, the invariant phase $\phi := k \cdot x = \omega t^-$ with light-front
time $t^- = x_0 - x_\parallel$ in a coordinate system where $\vec k \parallel \vec e_z$.  
Due to the high symmetry, the Dirac equation can be solved exactly, delivering
the Volkov solution, which depends trivially on three components of space-time and, generally,
highly nontrivial on $\phi$ or $t^-$ via the background. For a laser pulse, the background has finite
support in the variables $\phi$ and $t^-$, and the details of the temporal structure shape the final phase-space
distribution in a distinct manner. The limiting case of an infinitely long-acting external field is
called a monochromatic field. It is to be contrasted with the sandwich field, where - like in the
situation of a passing gravitational wave - an observer sees the vacuum, followed by the pulse,
and is left afterward in the vacuum again, irrespective of memory effects imparted on test particles.

It is often taken for guaranteed that the weak-field limit of the background field (which is a classical field)
and the monochromatic limit (i.e.\ infinite support at constant strength) yields the standard 
perturbative QED results obtained via Feynman diagrams in momentum space. The role of
loops is less obvious in that correspondence. A particular situation is when considering a short weak 
(classical background) field pulse: A straightforward treatment by perturbative QED via Feynman diagrams
does not catch the features caused by the higher/lower Fourier components of the weak field,
see \cite{HernandezAcosta:2019vok}.
In particular, \cite{Ilderton:2020rgk} points out that, to preserve Ward identities, one has to add
to the Furry-picture position-space Feynman diagrams some terms ``by hand".
It happens that a concise formulation in momentum space\footnote{
The nontrivial relation of position-space Feynman rules and momentum-space Feynman rules
is phrased in \cite{Agarwal:2021ais} as follows.
``When deriving the momentum-space Feynman rules, 
we have formally integrated over the position of the photon emission
vertex in spacetime, under the assumption that emission at long distances should be sufficiently
suppressed. Unfortunately, it is not, a consequence of the fact that QED (like all unbroken
gauge theories) has long-range interactions."}
provides, in a clear manner, 
these necessary terms, while their origin and relation to gauge invariance are obscured in position space. 
 
It is the goal of the present paper to dilate on the strong-field QED Furry-picture in the momentum space.
Our approach has been outlined in \cite{Acosta:2020dud}.
The key is to employ Ritus matrices \cite{Ritus85}
and to accumulate all dependencies on the external classical field in the dressed vertex
\cite{Meuren:2013oya}, while keeping vacuum photon and Fermion lines for propagators and
$in$/$out$ states. Our presentation enables easy access to the systematic expansion 
of amplitudes and probabilities/cross sections in powers of the classical laser intensity parameter $a_0$. 
In the lowest order of small $a_0$ we obtain ``pulsed perturbative QED" which accounts for
temporal pulse shape effects \cite{Acosta:2020dud,HernandezAcosta:2019vok};
the very special case of a monochromatic external classical field recovers standard perturbative QED. 

Our paper is organized as follows. In Section \ref{sect:Feynman} we present the formal development
of the momentum-space Feynman rules with emphasis on implementing gauge invariance,
definition of the Fermion-Fermion-photon vertex and the graphical representation.
The monochromatic limit is considered too
and some remarks on soft factors are supplied as well.
Examples of applications are introduced in Section \ref{sect:Examples}.
The explication of the three-point amplitude,
that is for the one-vertex processes nonlinear Compton/Breit-Wheeler/one-photon annihilation,
is spelt out in Section \ref{sect:one_vertex} to demonstrate the path from our rules 
towards the basic formulation of already exhaustively analyzed phenomena.   
The two two-vertex processes and related four-point amplitudes 
are dealt with in Section \ref{sect:two_vertex} with some details
w.r.t.\ non-linear two-photon Compton and nonlinear M{\o}ller scatterings and crossing channels.
Special aspects of Oleinik resonances, singularity structures, and the monochromatic limit are uncovered here.
We conclude in Section \ref{sect:Summary}. Appendix \ref{sect:Ward_ID} deals in some detail
with the regularization to ensure gauge invariance.
  
\section{Feynman rules for Furry picture in momentum space}\label{sect:Feynman}

\subsection{Background field description}

The following considerations apply to Lorenzian null fields, i.e.\ the classical background field has the structure
$\vec E = - \partial_t \vec A$ (electric field) and $\vec B = \vec \nabla \times \vec A$ (magnetic field) 
with four-potential $A^\mu = (A^0, \vec A)$
in Lorenz ($\partial_\mu A^\mu = 0$) \& Weyl ($A^0 = 0$) gauge
\begin{equation} \label{eq:Amu}
A^\mu = \frac{m}{\vert e \vert } a_0 g(\phi, \Delta \phi)
[ \epsilon_1^\mu \cos (\xi) \cos (\phi + \phi_{\mathrm{CEP}}) +   
   \epsilon_2^\mu \sin( \xi) \sin (\phi + \phi_{\mathrm{CEP}})        ] , 
\end{equation}
where, in units with $c = \hbar = 1$, $m$ and $e$ denote the electron's rest mass and electric charge,
$\epsilon_{1,2}^\mu$ refer to the laser's polarization vector
(e.g.\ $\epsilon_1^\mu = (0, 1 ,0,0)$ and $\epsilon_2^\mu = (0, 0, 1, 0)$
in a reference frame where $k^\mu = \omega (1, 0, 0, 1)$),   
and the carrier envelope phase reads $\phi_{\mathrm{CEP}}$.
Side conditions specify further this model of a laser pulse in plane-wave approximation:
\begin{eqnarray}
k^2 &:=& k \cdot k =0 \quad \mbox{(null field)},
\quad
k \cdot \epsilon_{1,2} = 0 \quad \mbox{(transversality)}, 
\quad \epsilon_i  \cdot \epsilon_j = - \delta_{ij}, \\
\xi &=& 
\left\{
\begin{array}{l}
0 \, \, \mbox{or} \, \frac{\pi}{2} \quad \mbox{(lin.\ polarization),} \\
\pm \frac{\pi}{4} \quad \mbox{(circ.\ polarization),}\\
\end{array} \right. 
\end{eqnarray}
or elliptic polarization for other values of $\xi$.
$g$ denotes the pulse shape function or envelope for a shortcut with \( \Delta \phi \) as the pulse width parameter.
Scalar products of four-vectors are henceforth noted as dot products.

\subsection{Dressed vertex decomposition} \label{dressed_vertex_decomposition}

The dressed vertex is defined by \cite{Meuren:2013oya}
\begin{eqnarray}
\Delta^\mu &:=& \int d^4 x \bar E_{p'} (x) (-i e \gamma^\mu ) E_p (x) e^{i k' \cdot x} \\
 &=& - \frac{i e}{2 \pi} \int d \ell \, \Gamma^\mu (\ell, p, p' \vert k) 
(2 \pi )^4 \delta^{(4)} (p + \ell k - p' - k'))
\end{eqnarray}
where $p$ ($p'$) is the in- (out-) going Fermion four-momentum, and the
outgoing photon four-vector is denoted by $k'$. By inserting the Ritus matrices, e.g.\
$E_p = ( 1 + e \frac{\slashed{k} \slashed{A}}{2 k\cdot p}) \exp\{ i S_p (x)\}$
with the Hamilton-Jacobi action
$S_p(x) = - p \cdot x - \frac{1}{2 k \cdot p} \int_{\phi_0}^{\phi = k \cdot x} d \phi'
[2 e p \cdot A(\phi') - e^2 A^2 (\phi')]$, the second line follows, defining the dressed vertex function:
\begin{align} \label{eq:def_vertex}
\Gamma^\mu (\ell, p, p' \vert k) &:= \int d \phi \, 
\left(1 - e \frac{\slashed{k} \slashed A}{2 k \cdot p'} \right)
\gamma^\mu
\left(1 + e \frac{\slashed{k} \slashed A}{2 k \cdot p}\right)
\exp \{ S_{p'} - S_p + p' \cdot x  - p \cdot x \} \\
 &= \gamma^\mu B_0 (\ell) + \Gamma_1^{\mu \nu} B_{1 \nu} (\ell) + \Gamma_2^\mu B_2 (\ell) .
\label{eq:vertex}
\end{align}   
Note (i) the crucial ``photon number parameter" $\ell$ as Fourier conjugate of the phase $\phi$
and (ii) the decomposition into elementary vertices $\{\gamma^\mu, \Gamma_1^{\mu \nu}, \Gamma_2^\mu \}$,
depending on $\{p, p', k \}$,  
\begin{align} 
\gamma^\mu &: \quad \mbox{Dirac matrix obeying} \quad [\gamma_\mu, \gamma_\nu]_+ = 2 g_{\mu \nu}, \\
\Gamma_1^{\mu \nu} &= e \left(
\frac{\gamma^\mu \slashed{k} \gamma^\nu}{2 k \cdot p} + 
\frac{\gamma^\nu \slashed{k} \gamma^\mu}{2 k \cdot p'} \right) , \label{eq:vertex1}
\\ 
\Gamma_2^\mu &= - e^2 \frac{\slashed{k} k^\mu}{2 k \cdot p k \cdot p'} , \label{eq:vertex2}
\end{align}
and phase integrals $\{ B_0, B_1^\mu, B_2 \}$, depending on $\ell$ as well
\begin{align}
B_0 &= \int_{- \infty}^{\infty} d \phi \exp\{ i \ell \phi + i G(\phi) \} , \label{eq:B0}  \\
B_1^\mu &= \int_{- \infty}^{\infty} d \phi \exp\{ i \ell \phi + i G(\phi) \} A^\mu (\phi) , \label{eq:B1}
\\ 
B_2 &= \int_{- \infty}^{\infty} d \phi \exp\{ i \ell \phi + i G(\phi) \} A^2 (\phi) . \label{eq:B2}
\end{align} 
The function $G$ reads
\begin{equation} \label{eq:G}
G(\phi, \phi_0) = \alpha_1^\mu \int_{\phi_0}^\phi d \phi' \, A_\mu (\phi') +
\alpha_2 \int_{\phi_0}^\phi d \phi' \, A^2 (\phi'),
\end{equation}
where $\phi_0 \to -\infty $ is a useful choice for pulses and 
\begin{eqnarray} \label{eq:alpha12}
\alpha_1^\mu = e \left( \frac{p'{}^\mu}{k \cdot p'}  - \frac{p^\mu}{k \cdot p} \right) ,
\quad
\alpha_2 = - e^2 \left( \frac{1}{k \cdot p'}  - \frac{1}{k \cdot p} \right) .
\end{eqnarray}
The phase integral (\ref{eq:B0}) in Eq.~(\ref{eq:vertex}) needs a regularization 
which takes care of the Ward identity $k' \cdot \Gamma = 0$.
This is accomplished by (see Appendix \ref{sect:Ward_ID})
\begin{equation} \label{eq:B0_reg}
B_0 (\ell) = \pi  {\cal G}  \delta(\ell) - {\cal P} \left( \frac{\alpha_1^\mu B_{1 \mu}}{\ell} +
 \frac{\alpha_2 B_2}{\ell} \right) ,
\end{equation}
where the instruction ${\cal P}$ means taking Cauchy's principle value and
\begin{equation} \label{eq:Gpm}
{\cal G} = \exp\{ i G^+ \} + \exp\{ i G^-\}, 
\quad
G^\pm := \lim_{\phi \to \infty} G(\pm \phi).
\end{equation}
Eventually, the dressed vertex function (\ref{eq:def_vertex}, \ref{eq:vertex}) 
is decomposed as
\begin{align}
\Gamma^\mu (\ell) &= \Gamma_{\mathrm{div}}^\mu (\ell) + \Gamma_{\mathrm{reg}}^\mu (\ell),  \label{eq:Gamma_decomp} \\
\Gamma_{\mathrm{div}}^\mu (\ell) &:= \gamma^\mu \pi {\cal G} \delta(\ell), \label{eq:Gamma_div}\\
\Gamma_{\mathrm{reg}}^\mu (\ell) &:=
\left( \Gamma_1^{\mu \nu}(\ell) - {\cal P} \frac{\gamma^\mu \alpha_1^\nu}{\ell} \right) B_{1 \nu} (\ell)+
\left( \Gamma_2^\mu - {\cal P} \frac{\gamma^\mu \alpha_2}{\ell} \right) B_2 (\ell). \label{eq:Gamma_reg}
\end{align}
We have suppressed the pertinent arguments $p, p', k$ in all functions, but highlighted the $\ell$ dependence. 
The term labeled by ``div" could be named ``gauge invariance restoration term", 
since it emerges just from that requirement.

The decomposition of the vertex function implies the following modification
of standard Feynman rules in momentum space:
use $\int \frac{d \ell}{2 \pi} (-ie) \Gamma_\mu (\ell)$ for the Fermion-Fermion-photon vertex, 
instead of $-ie \gamma_\mu$, and integrate over internal momenta.  
While a $N$-vertex Furry-picture diagram in position space involves $N$ space-time integrals
(cf.\ \cite{Fedotov:2022ely} of how to process the expressions),
in momentum space one meets $N$ integrations over the respective vertex-attached ``photon number parameter" $\ell$.

In a follow-up paper, we show in more detail that, in leading order of a series expansion in powers
of $a_0$, the standard momentum Feynman rules are recovered and explicate the NLO terms.  

\subsection{Expansion in powers of $\mathbf{a_0}$}

One benefit of our momentum space formulation is the possibility 
of a straightforward series expansion of the amplitude in powers of $a_0$.
The temporal pulse shape is imprinted transparently in the weak-field limit.
To begin with, we introduce for the bookkeeping of powers of $a_0$ the tilde notation: 
every quantity with a tilde is free of any dependence on $a_0$, 
e.g.\ $A^\mu (a_0, \phi) = a_0 \tilde A^\mu(\phi)$, which implies
\begin{align} \label{eq:defG}
G &= a_0 \, \alpha_1^\mu \int_{\phi_0}^\phi d \phi' \, \tilde A_\mu (\phi') +
a_0^2 \, \alpha_2 \int_{\phi_0}^\phi d \phi' \, \tilde A^2 (\phi')
\equiv a_0 \tilde A_1 + a_0^2 \tilde A_2 , \\
\exp\{ i G \} &= \sum_{N = 0}^\infty a_0^N \tilde G_N \equiv
\sum_{N = 0}^\infty a_0^N \sum_{(m, n)} \tilde G_{m n}\vert_{m + 2n = N}, \quad
\tilde G_{m n} := \frac{i^{m + 2 n}}{m! n!} \tilde A_1^m \tilde A_2^n ,
\end{align}
when using the abbreviations $\tilde A_1 := \alpha_1^\mu \int_{\phi_0}^\phi d\phi' \tilde A_\mu (\phi')$ and
$\tilde A_2 := \alpha_2 \int_{\phi_0}^\phi d\phi' \tilde A^2 (\phi')$. The quantities 
$\tilde G_N = \sum_{m, n \in {\cal N}(m,n)} \tilde G_{m n}$ use the re-indexing
with index set ${\cal N}(m,n) := \{ (m, n) \in \mathbb{N}^2 \vert m + 2 n = \mathbb{N} \}$,
e.g.\ $\{ (0, 0) \}$ for $N = 0$,
$ \{ (1, 0 )\}$ for $N = 1$, and 
$\{(2,0), (0,1)\}$ for $N = 2$.

Analogously, the phase integrals (\ref{eq:B1}, \ref{eq:B2}) become sums of Fourier transforms: 
\begin{eqnarray}
B_1^\mu &=& \sum_{N = 0}^\infty a_0^{1 + N} \tilde B_{1, N}^\mu, \quad
\tilde B_{1, N}^\mu := \int_{- \infty}^\infty d \phi e^{i \ell \phi} \tilde G_N \tilde A^\mu , \label{eq:B1_a0_expand} \\
B_2 &=& \sum_{N = 0}^\infty a_0^{2 + N} \tilde B_{2, N}, \quad
\tilde B_{2, N} := \int_{- \infty}^\infty d \phi e^{i \ell \phi} \tilde G_N \tilde A^2  .
\end{eqnarray}
Proceeding in such a manner we represent the vertex functions (\ref{eq:Gamma_div}, \ref{eq:Gamma_reg}) as
\begin{align}
\Gamma_{\mathrm{div}}^\mu &= \sum_{N = 0}^\infty a_0^N \tilde \Gamma_{div \, N}^\mu,
\quad 
\tilde \Gamma_{\mathrm{div} \, N}^\mu =
\pi \gamma^\mu \delta (\ell) (\tilde G_N^+ + \tilde G_N^-),  \label{eq:18}\\
\Gamma_{\mathrm{reg}}^\mu &= \sum_{N = 1}^\infty a_0^N \tilde \Gamma_{\mathrm{reg} \, N}^\mu,
\quad
\tilde \Gamma_{\mathrm{reg} \, N}^\mu = 
\left( \Gamma_{1 \nu}^\mu - {\cal P} \frac{\gamma^\mu \alpha_{1 \nu}}{\ell} \right)
\tilde B_{1, N - 1}^\nu  \nonumber \\
&  \hspace*{4.5cm}+ \Theta(N - 1)
\left( \Gamma_2^\mu - {\cal P} \frac{\gamma^\mu \alpha_2}{\ell} \right)
\tilde B_{2, N - 2} , \label{eq:19}
\end{align}
where the Heaviside distribution ensures to take only $N > 1$ contributions in the last term.
We note (i) 
$\tilde \Gamma_{\mathrm{div} \, N=0}^\mu = 2 \pi \gamma^\mu \delta(\ell)$
due to $\tilde G_{N = 0}^\pm = 1$;
$\delta(\ell)$ enforces $\ell = 0$ meaning that this vertex part
does not exchange energy-momentum with the external field, and
(ii) the $N = 1$ contribution to $\Gamma_{\mathrm{reg}}^\mu$ is solely related to 
$\tilde B_{1}^\mu = \int_{- \infty}^\infty d \phi \, e^{i \ell \phi} \tilde A^\mu (\phi)$,
i.e.\ the Fourier transform of the background field; it is the only linear contribution
(in $a_0$ and the background field), while contributions $\propto a_0^N$ with $N > 1$
are nonlinear in the background field.
 
\subsection{Graphical representations}

The above vertex function decomposition facilitates the following graphical representation
with the $a_0$ expansion in the bottom lines:
\begin{align}
\underbrace{\includegraphics[width=0.28\textwidth,valign=c]{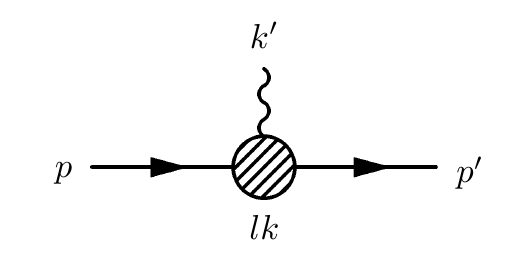}}_{\Gamma^\mu}
 =
\underbrace{\includegraphics[width=0.28\textwidth,valign=c]{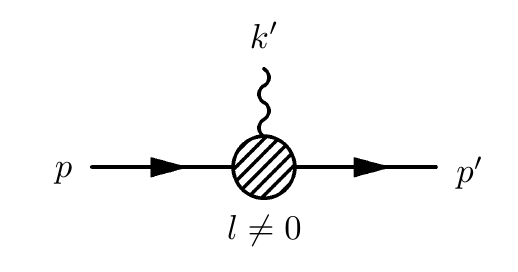}}_{\Gamma_{\mathrm{reg}}^\mu, \,
\text{Eq.}~(\ref{eq:Gamma_reg}) }
&  \, + \,
\underbrace{\includegraphics[width=0.28\textwidth,valign=c]{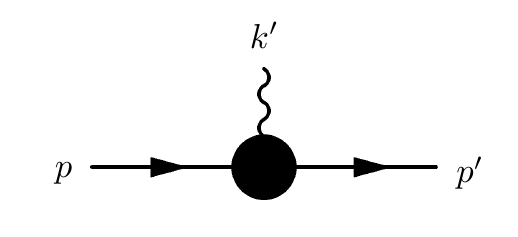}}_{\Gamma_{\mathrm{div}}^\mu, \,
\text{Eq.}~(\ref{eq:Gamma_div}) }\\
= \sum_{N = 1}^\infty a_0^N \hspace*{-6mm}
\underbrace{\includegraphics[width=0.28\textwidth,valign=c]{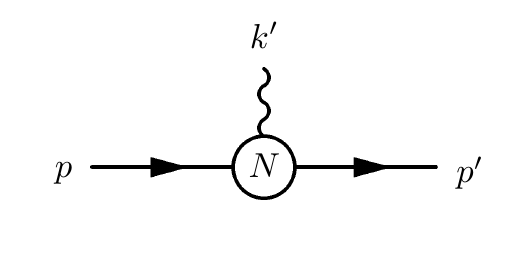}}_ {\tilde \Gamma_{\mathrm{reg} \, N}^\mu, \, \text{Eq.}~(\ref{eq:19}) }
& \, + \, \sum_{N = 0}^\infty a_0^N \hspace*{-6mm}
\underbrace{\includegraphics[width=0.28\textwidth,valign=c]{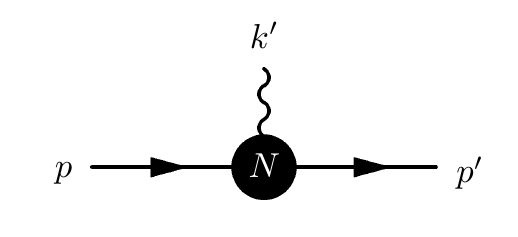}}_{ \tilde \Gamma_{\mathrm{div} \, N}^\mu, \, \text{Eq.}~(\ref{eq:18}) } .
\label{diag:Nexpansion}
\end{align}

Each expression must be supplemented by a factor of $-i e \delta^{(4)}(p + \ell k -p' - k')$.
The pertinent arguments $\ell, p, p', k$ are not displayed here.
Note that the expanded vertex functions in (\ref{diag:Nexpansion})
are $\propto a_0^N$, thus allowing for easy bookkeeping.

\subsection{Monochromatic limit: \( \Delta \phi \to \infty \)}

The above formalism is devised to include arbitrary shape functions $g(\phi, \Delta \phi)$.
For specified (explicitly given) temporal pulse structures, $G(\phi)$ can be explicated and
the phase integrals $B_0$, $B_1^\mu$ and $B_2$
can be executed to arrive at special functions
(cf.\  \cite{Seipt:2016rtk} for such an example w.r.t.\ nonlinear Compton). 
A very special case is the monochromatic plane-wave background field. Formally, \( \Delta \phi \to \infty \) and
the envelope function in Eq.~(\ref{eq:Amu}) obeys $g(\phi, \Delta \phi) = 1$
for all $\phi \in \mathbb{R}$. 
Setting $\phi_0 = 0$ is a suitable choice for monochromatic plane-wave backgrounds, since then
the initial condition is known due to $A^\mu (\phi_0) = a \epsilon_1^\mu \cos \xi$ and 
the nonlinear phase defined in (\ref{eq:G}) is finite for finite values of $\phi$.
The distinction of $\Gamma_{\mathrm{div}}^\mu$ and $\Gamma_{\mathrm{reg}}^\mu$ is no longer suitable,
since also $B_0$, $B_1^\mu$ and $B_2$ in $\Gamma_{\mathrm{reg}}^\mu$ contribute to the so-called
$\delta$-comb, which refers to individual harmonics.
The carrier-envelope phase $\phi_{\mathrm{CEP}}$ is irrelevant and can be skipped.  

Under these prepositions, Eq.~(\ref{eq:defG}) yields
\begin{align}
G(\phi, \phi_0 = 0) = 
-\vert \alpha_- \vert \left[ \sin( \phi + \Theta) - \sin( \Theta) \right] 
- \beta \left[ \frac12\cos(2 \xi) \sin(2 \phi) + \phi \right] ,
\end{align} 
where \( \alpha_- := a \, \alpha_{1} \cdot\epsilon_{-}\) 
is represented as  $\alpha_-= \vert \alpha_- \vert \mathrm{e}^{ i \Theta}$, 
and $a \equiv a_0 m/\vert e \vert$, \( \beta := \frac12 a^2 \alpha_2 \). 
Insertion into Eqs.~(\ref{eq:B0}, \ref{eq:B1}, \ref{eq:B2}) results in
\begin{align}
B_0(\ell) &= 
2 \pi \exp\{ i\vert \alpha_-\vert \sin \Theta\}  \sum_{n=-\infty}^\infty 
\, \delta\left( \ell - \beta- n\right) C_{n} , \\
B_{\pm}(\ell) &=  
2 \pi \exp\{i\vert \alpha_-\vert \sin \Theta \} \sum_{n=-\infty}^\infty 
\, \delta\left( \ell - \beta - n\right) C_{n \mp 1} ,  \label{eq:Bpm_IWA}\\
B_2(\ell) &=  
2 \pi \exp\{i\vert \alpha_- \vert \sin \Theta\} \sum_{n=-\infty}^\infty 
\, \delta\left( \ell - \beta - n \right) \left[C_{n} + \frac12 \cos (2 \xi) (C_{n+2} + C_{n-2})\right] 
\end{align}
and allows the representation of the fully dressed vertex as
\begin{align}
\Gamma^\mu(\ell ) &=  
\sum_{n=-\infty}^\infty \delta\left( \ell - \beta- n \right) \Gamma_{\mathrm{IPW,n}}^\mu\\
\Gamma_{\mathrm{IPW,n}}^\mu & := \label{eq:Gamma_IPW}
2 \pi \exp\{i \vert \alpha_- \vert \sin \Theta\} \\
&\times\left\{\gamma^ \mu C_n 
+ \frac12 a \left[\Gamma_{+}^\mu C_{n - 1} + \Gamma_{-}^\mu C_{n + 1} \right] 
+ \Gamma_2^ \mu\left[C_{n} 
+ \frac12 \cos (2 \xi) (C_{n+2} + C_{n-2})\right] \right\}, \nonumber
\end{align}
where we use the abbreviation \( \Gamma_\pm^ \mu := \epsilon_{-, \nu} \Gamma_1^{ \mu \nu}\);
cf.\ Eqs.~(\ref{eq:Bpm}, \ref{eq:Bpmeps}) below for the definition of $\epsilon_\pm^\mu$ and
the decomposition of $B_1^\mu$ into $B_\pm$. 
To make some intermediate steps more obvious, we note
\begin{align}
B_\pm(\ell) &:= \int_{-\infty}^\infty  \mathrm{d} \phi\, \exp\{ i(\ell \pm 1) \phi + iG(\phi) \} \\
&=  
\exp\{i \vert \alpha_- \vert \sin \Theta \}
\int_{-\infty}^\infty  \mathrm{d} \phi \, 
\exp\{ i\left( \ell \pm 1 - \beta \right) \phi - i\vert \alpha_- \vert \sin( \phi + \Theta)  \nonumber \\
& \hspace*{8.35cm} 
- \frac{i}{2} \beta \cos(2 \xi) \sin(2 \phi) \} . \label{eq:Bpm_expl}
\end{align}
Using the Jacobi-Anger expansion of the non-Fourier part one gets
\begin{align}
& \exp\{ - i \vert \alpha_- \vert \sin( \phi + \Theta) - \frac12  i \beta \cos(2 \xi) \sin(2 \phi)\} 
= \sum_{n=-\infty}^{\infty} C_n \mathrm{e}^{-i n \phi}, \\
& C_n = \sum_{s = -\infty}^\infty J_{n - 2s}( \vert \alpha_-\vert ) J_s \left(\frac12 \beta \cos(2 \xi) \right) 
\mathrm{e}^{-i(n-2s) \Theta},
\end{align}
where $J_n$ stand for Bessel functions of the first kind.
The phase integral (\ref{eq:Bpm_expl}) reads then
\begin{align}
B_{\pm}(\ell) &=  \exp\{i\vert \alpha_- \vert \sin \Theta \} 
\sum_{n=-\infty}^\infty \sum_{s = -\infty}^\infty J_{n - 2s}(\vert \alpha_- \vert ) \,
J_s \left( \frac12 \beta \cos(2 \xi)\right) \mathrm{e}^{-i(n-2s) \Theta}\\
&\times \int_{-\infty}^\infty  \mathrm{d} \phi\, \exp\{ i\left(\ell \pm 1 - \beta - n\right) \phi \} ,
\end{align}
and the index shift $n \to n \pm 1$ yields Eq.~(\ref{eq:Bpm_IWA}).

\subsection{Recovery of Feynman rules of perturbative QED} \label{recovery_Feynman}

The relation to standard perturbative QED Feynman rules has two facets.

(i) Get the Fermion-Fermion-photon vertex from the dressed vertex by inspecting Eq.~(\ref{diag:Nexpansion})
and noting that, at $a_0 \to 0$ and for off-shell legs, $\mathcal{G} \to 2 \pi$ and the leading order term in the sum
$\sum_{N = 1}^\infty \cdots \to a_0 \to 0$, i.e.\
\( \includegraphics[width=0.13\textwidth,valign=c]{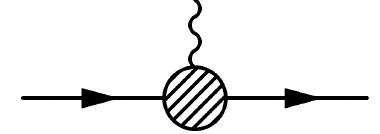} \to 0  \)
and \( \includegraphics[width=0.13\textwidth,valign=c]{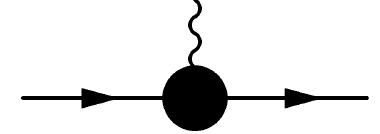} \to 
\includegraphics[width=0.13\textwidth,valign=c]{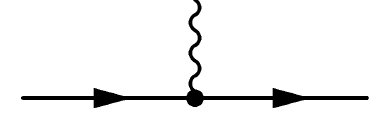} \).

(ii) However, in the case of a monochromatic background and for on-shell legs
(e.g.\ for Compton),
the leading-order terms in the $a_0$ expansion stem from the vertex
\( \includegraphics[width=0.12\textwidth,valign=c]{sf_mom_vertex_fin_nolabel} = \sum_{N = 1}^\infty a_0^N 
\includegraphics[width=0.12\textwidth,valign=c]{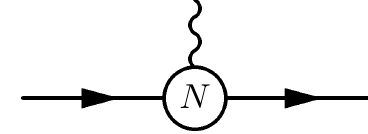}
 \) since (for on-shell amplitude) \includegraphics[width=0.12\textwidth,valign=c]{sf_mom_vertex_inf_nolabel} $= 0$.
The powers of $a_0$ are expelled off the regularized vertex, which becomes   
\includegraphics[width=0.12\textwidth,valign=c]{sf_mom_vertex_fin_Nth_nolabel} $\to$
\includegraphics[width=0.12\textwidth, valign=c]{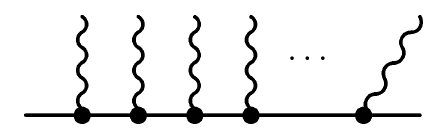} 
with $N$ incoming laser photon lines 
\includegraphics[width=0.02\textwidth,valign=c]{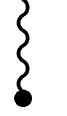} in all permutations.

On the level of cross-section,
$d \sigma / d \Omega' \, d \omega' = \mathcal{I}_\gamma^{-1} \sum \vert \mathcal{M} \vert^2 d \Phi'$,
with $\mathcal{I}_\gamma \propto a_0^2$. Therefore, for $a_0 \to 0$, only the term
\( \mathcal{M} \to a_0^{N=1} ( \includegraphics[width=0.12\textwidth,valign=c]{sf_mom_vertex_fin_Nth_nolabel} )
\overset{\Delta \phi \to \infty}{=}
\includegraphics[width=0.12\textwidth, valign=c]{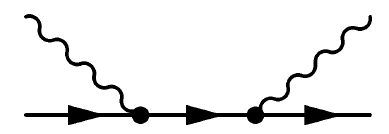} +
\includegraphics[width=0.12\textwidth, valign=c]{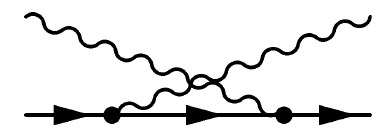} 
 \)
survives, making the cross-section $\propto \sum \vert \mathcal{M} \vert^2 
\delta (\omega' - \frac{\omega}{1 + \frac{\omega'}{m} (1 - \cos \Theta' )} )$, where, upon executing the
spin and polarization sums, $\sum$, the Klein-Nishina cross section is recovered; the energy-momentum balance
via the delta distribution arises from phase space integration, $d \Phi'$, and only for $\Delta \phi \to \infty$.

\subsection{Soft photons: $k' \to 0$}\label{sec:soft_emission}

Lowest-order soft theorems\footnote{
For the relation of soft-photon theorem and asymptotic symmetry (Ward identity)
and memory effect within the infrared triangle, see \cite{Strominger:2017zoo} and citations therein.}
allow for factorizing amplitudes as
${\cal M} = {\cal M}_{hard} \times S$, where $S$ is the soft factor accounting for the emission of a soft photon
and ${\cal M}_{hard}$ is the amplitude of hard interaction.
Focusing first on the soft-photon emission off an external leg, e.g.\ an incoming electron,
the corresponding matrix element reads
\begin{align}
{\cal M} &= 
\includegraphics[width=0.28\textwidth,valign=c]{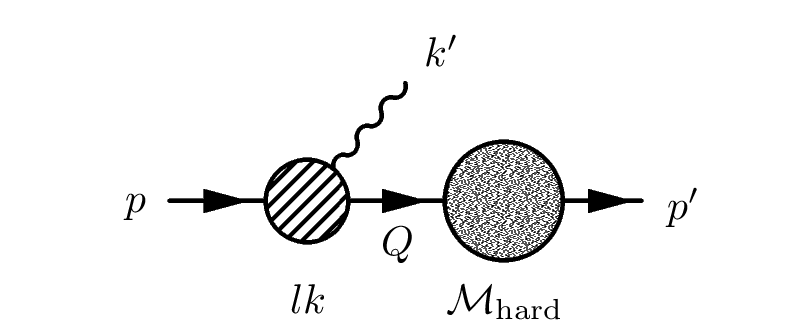}
+  
\includegraphics[width=0.28\textwidth,valign=c]{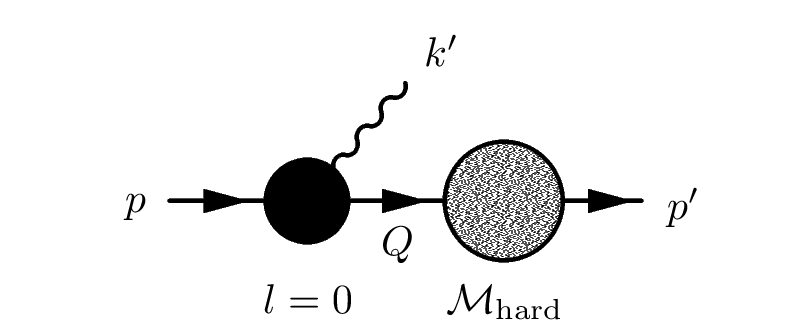} \\
&= - i e \int \frac{d \ell}{2 \pi} \left[ \bar u (p') 
{\cal M}_{hard}(p', p + \ell k - k')  \, S_F(p + \ell k -k') \Gamma^\mu(\ell, p + \ell k - k') ) {\epsilon_\mu^*}' (k') \, u(p) \right] ,
\label{eq:soft_photon}
\end{align} 
where ${\cal M}_{hard}$ is part of the matrix element which emerges from the interaction of the electron with other
non-soft particles and ${\epsilon_\mu}' (k')$ denotes the polarization of the emitted soft photon.
Since $k'$ is small compared with the electron momenta, one has 
${\cal M}_{hard} (p', p + \ell k - k') \approx {\cal M}_{hard} (p', p + \ell k)$. Furthermore, we observe
$\Gamma^\mu (\ell, p, p+ \ell k - k') \equiv  \Gamma^\mu (\ell, p, p - k')$ since the out-going momentum
$p'$ appears only in the product $p' \cdot k$ in $\Gamma^\mu (\ell, p, p')$.
Inserting the decomposition (\ref{eq:Gamma_decomp}) in the matrix element (\ref{eq:soft_photon}), 
the infrared behavior of $k'$ must be examined for two parts:\\
(i) The gauge-restoration part $\Gamma^\mu_{\mathrm{div}}$:
The corresponding matrix element becomes 
$\mathcal{M}_{\mathrm{div}} =  - i e \int \frac{d \ell}{2 \pi} [ \bar u (p') \mathcal{M}_{hard} (p', p + \ell k) S_F(p + \ell k - k')
\pi \mathcal{G} (p, p- k') \delta(\ell) \gamma^\mu {\epsilon_\mu^*}' (k') u(p)]$,
where the $\ell$ integration is executed by employing the $\delta$-distribution leading to $\ell = 0$.
Considering the nonlinear Volkov phase $G$ defined in \eqref{eq:G}, we have 
$\lim_{k' \to 0} G(\phi, p, p- k') =  G(\phi, p, p) \equiv 0$, which implies
$\lim_{k' \to 0} \mathcal{G}(p, p- k') = 2$.  That means this part of the matrix element has the same infrared behavior
as ordinary one-photon bremsstrahlung (cf.\  sections 6.1 in \cite{Peskin:1995ev}
and 13.5 in \cite{Weinberg:1995mt}):
$\mathcal{M}_{\mathrm{div}} \approx - \bar u(p) \mathcal{M}_{hard}(p',p) u(p) \left[e\frac{p \cdot \epsilon'}{k' \cdot p}\right]$,
where we used $S_F(p - k') \gamma^\mu {\epsilon_\mu^*}' u(p) \to - i \frac{p \cdot \epsilon'}{p \cdot k'}$
for $k' \to 0$.  \\
(ii) The regularized finite part $\Gamma^\mu_{\mathrm{reg}}$:
Considering the leading order in powers of $a_0$ and the monochromatic limit, $\Delta \phi \to \infty$,
the diagram (\ref{eq:soft_photon}) restores the standard perturbative case:
\begin{align}
\includegraphics[width=0.22\textwidth,valign=c]{soft-gamma-radiation.png} 
\xrightarrow[\lim a_0 \to 0]{\lim \Delta \phi \to \infty}
\underbrace{\includegraphics[width=0.22\textwidth,valign=c]{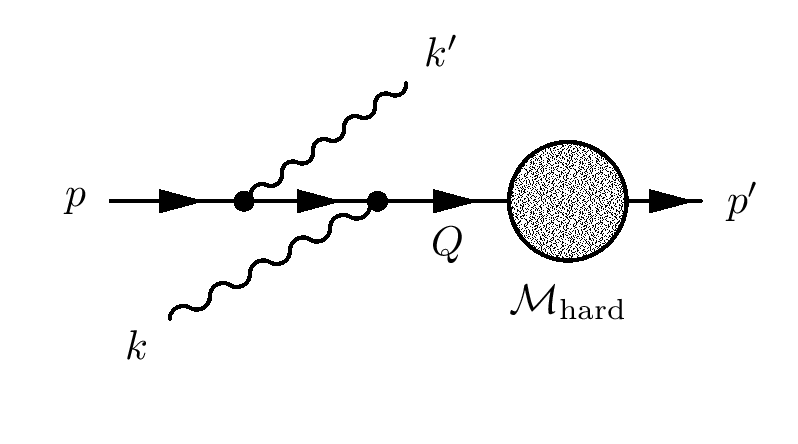}}_{\mathcal{M}_{\mathrm{reg}}^{(1)}}
+
\underbrace{\includegraphics[width=0.22\textwidth,valign=c]{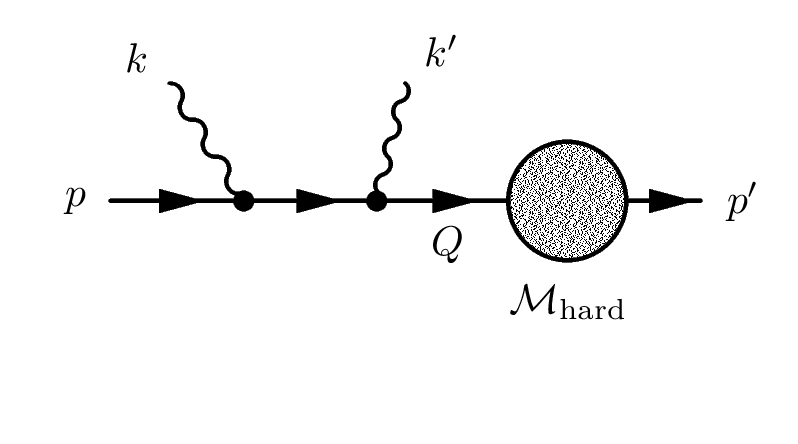}}_{\mathcal{M}_{\mathrm{reg}}^{(2)}}
\end{align}
Therefore, we have a special case of a sub-leading Low theorem, where one part of the matrix element, 
$\mathcal{M}_{\mathrm{reg}}^{(1)}$, has an infrared pole and the other term, $\mathcal{M}_{\mathrm{reg}}^{(2)}$, does not.
The elaboration of  the general version of $\mathcal{M}_{reg}^{(1, 2)}$ is relegated to separate work.

\subsection{Soft background field: \( k\to 0 \)}\label{sec:soft_background}

Similarly to Section \ref{sec:soft_emission}, the soft interaction with the background field refers to the limit \( k\to 0 \), which is equivalent to \( \omega \to 0 \), where we set \( k ^\mu  = \omega n ^\mu \) with the normalized reference momentum \( n ^\mu  \). Consequently, this soft limit needs to be examined separately for the two parts of the dressed vertex. 

(i) \emph{Regularized part \( \Gamma_{\mathrm{reg}}^ \mu \)} --
 Considering the elementary vertices defined in Eqs.~\eqref{eq:vertex1} and \eqref{eq:vertex2}, 
we find \( \Gamma_1^{ \mu \nu} =  e \left(\frac{\gamma^\mu \slashed{n} \gamma^\nu}{2 n \cdot p} + \frac{\gamma^\nu \slashed{n} \gamma^\mu}{2 n \cdot p'} \right) \) and \( \Gamma_2^\mu = - e^2 \frac{\slashed{n} n^\mu}{2\, n \cdot p\, n \cdot p'} \), which are both finite in the limit \( \omega\to 0 \). On the contrary, the kinematic factors \( \alpha_1 ^\mu , \alpha_2 \) defined in \eqref{eq:alpha12} have the typical form of Weinberg's soft factors \cite{Weinberg:1995mt}, hence a divergence in the soft limit. Therefore, for \( \omega\to 0 \), we have \(  \Gamma_1^{ \mu \nu}  \ll \frac{ \gamma ^\mu \alpha ^\nu _1}{\ell}\) and  \(  \Gamma_2^{ \mu}  \ll \frac{ \gamma ^\mu \alpha_2}{\ell}\) for all finite \( \ell \). This leads to the soft limit of the finite part of the dressed vertex \eqref{eq:Gamma_reg}: \( \lim_{ \omega\to 0} \Gamma_{\mathrm{reg}}^\mu (\ell) = \gamma ^\mu S^{soft}_{\mathrm{reg}}(\ell)\), where the regularized soft factor reads
\begin{align}
S^{soft}_{\mathrm{reg}}(\ell) = -\mathcal{P} \left[ \frac{ \alpha ^\nu _1}{\ell}B_{1 \nu}(\ell) + \frac{ \alpha_2}{\ell}B_2(\ell)\right]. 
\end{align}
Linearizing the phase integrals in \( a_0 \), i.e. considering only \( \tilde B_{1,1} ^\mu (\ell) \) given in \eqref{eq:B1_a0_expand} and dropping \( B_2 \sim a_0^2 \), this soft-factor results in \( S^{soft}_{\mathrm{reg}}(\ell) \to -\frac{ \alpha_1 ^\mu }{\ell} \int_{\infty}^{\infty}A _\mu (\phi) e^{i \ell \phi}  \), which recovers the soft-factor elaborated in \cite{Ilderton:2020rgk} for the case of Bhabha scattering. Furthermore, considering linear polarisation \( A ^\mu = a \epsilon ^\mu g(\phi, \Delta \phi) \cos \phi  \) and performing the simultaneous limit \( a_0\to 0 \) and \( \Delta \phi\to \infty \), the soft-factor reads \( S^{soft}_{\mathrm{reg}}(\ell) \to -a \alpha_1 ^\mu \epsilon = a e \left(\frac{p \epsilon}{k \cdot p} - \frac{p' \epsilon}{k \cdot p'} \right)\), which indeed is, up to a constant normalization, Weinberg's well-known soft-factor.

(ii) \emph{Gauge restoration part  \( \Gamma_{\mathrm{div}}^ \mu \) } -- First we note 
that the integrals appearing in the prefactor of the gauge restoration part \( \mathcal{G} \) 
defined in Eq.~\eqref{eq:Gpm},  \( \int_{ \phi_0}^{\pm\infty} A ^\mu (\phi')\, \mathrm{d} \phi' \) 
and \( \int_{ \phi_0}^{\pm\infty} A^2(\phi')\, \mathrm{d} \phi' \), 
are finite in the soft limit \( \omega\to0 \), whereas the exponents \( G_{\pm}\) of \( \mathcal{G} \) 
diverge due to the presence of the factors \( \alpha_1 ^\mu , \alpha_2 \). 
This means, the factor \( \mathcal{G} \) itself acts like a soft factor, which highly oscillates in the soft limit. 
However, considering the linearization of this soft factor in \( a_0 \), 
we find \( \mathcal{G}\to i \alpha_1 ^\mu \left[\int_{ \phi_0}^{+\infty} A _\mu (\phi')\, \mathrm{d} \phi' + \int_{ \phi_0}^{-\infty} A _\mu (\phi')\, \mathrm{d} \phi'\right]\), 
which again recovers the form known from Weinberg's soft factors, 
but this time with the integrated fields acting as polarization vectors.
\\\\
In summary, it can be stated that in both cases, soft photon emission and soft interaction with the background field, generalized versions of typical soft factors appear, which can be, in a suitable limit, connected to soft factors well-known from monochromatic QED. However, the cancellation of the soft factors shown here with higher-order vertex corrections, i.e. finding a generalized version of the Bloch-Nordsieck theorem, deserves separate work.

\section{Examples \& future applications}\label{sect:Examples}

The above formalism is ready for direct numerical applications.
Elements are Dirac spinors, Dirac matrices, the metric tensor,
momentum, and polarization four-vectors; 
Fermion and photons propagators are as in free-field and could be defined as stand-alone objects; 
most importantly, the nonlinear phase integrals encode solely the external field and
require some care and numerical optimization. 
For a given exclusive reaction, these elements are to be connected
by scalar and matrix products, thus delivering a few partial amplitudes 
(e.g.\ direct and exchange terms or the multitude of diagrams with the
same $out$-state)
to be summed up to one complex number -- the amplitude ${\cal M}$.
Its mod-square, $\vert {\cal M} \vert^2$, is to be garnished to arrive eventually 
at probability or cross-section which depend on spins, polarizations, and
invariants referring to the initial state including the background field 
and final phase space. Partial or complete integration over the final
phase space variables need often specially adapted procedures, 
while spin/polarization summations, if required, are straightforward. 
Handling of the $\delta$ distributions is analog to position space formulation:
All $\ell$ dependence is integrated out before squaring the amplitude,
and finally use $[(2 \pi)^3 \delta^{(3)} (p_i - p_f) ]^2 \to (2 \pi)^3 V \frac{p_i^0}{p_i^+} \delta^{(3)} (p_i - p_f)$.     
We refrain here from such specific numerologies but instead stress the need for an in-depth understanding 
of the essential dependencies and singular structures of ${\cal M}$,
as the the core of $\vert {\cal M} \vert^2$, prior to numerical evaluations.

To sketch applications of the presented formalism 
we (re)consider  one- and two-vertex processes related to 
three- and four-point amplitudes of
nonlinear (one- and two-photon) Compton and M{\o}ller scattering
processes.
The detailed application to nonlinear trident is relegated to an accompanying paper.
The following Section \ref{sect:one_vertex} recalls the one-vertex processes by demonstrating how the above rules
lead to the known matrix elements, in particular for nonlinear Compton with a few supplementing remarks on
nonlinear Breit-Wheeler. The next-to-one Section \ref{sect:two_vertex}
considers the two two-vertex processes with emphasis
on nonlinear two-photon Compton and nonlinear M{\o}ller scattering.

\section{One-vertex processes/three-point amplitude}\label{sect:one_vertex}

The matrix element of one-vertex processes has, symbolically, the structure
${\cal M} \sim J_L^\mu E_\mu(k)$ with current $J_L^\mu \sim \bar u_L(p') \Gamma^\mu u_L(p)$,
where the label ``$L$" is a reminder of the laser dressing of charged Fermions with wave functions
$u$ and its adjoints $\bar u$, and $E_\mu(k)$ stands for the photon 
(momentum $k$) wave function. Depending on the orientation of the four-momenta $p$, $p'$ and $k$,
it refers to the processes $e_L^- \to {e_L^-}' + \gamma$ (nonlinear Compton),
$\gamma \to e_L^- + e_L^+$ (nonlinear Breit-Wheeler) 
and $e_L^- + e_L^+ \to \gamma$ (nonlinear one-photon annihilation), which are interrelated by crossing symmetry.
Due to CPT invariance, $e_L^\pm \Rightarrow e_L^\mp$ applies. 

\subsection{Nonlinear Compton}\label{sect:nlC}

The matrix element for nonlinear Compton (nlC) scattering reads
\begin{align} \label{eq:nlC_0}
M_{\mathrm{nlC}} = \int \frac{d \ell}{2 \pi} \delta^{(4)} (p + \ell k - p' - k')
\left[\bar u(p') (- i e) \Gamma^\mu (\ell, p, p', k) {\epsilon_\mu^*}' (k') u(p) \right] ,
\end{align} 
where ${\epsilon_\mu^*}' (k')$ stands for the polarization four-vector of the outgoing photon,
and we mark tied Fermion lines by $[ \cdots ]$.
The vertex decomposition (\ref{eq:Gamma_decomp}) facilitates two contributions,
$M_{\mathrm{nlC}}^{\mathrm{div}}$ and $M_{\mathrm{nlC}}^{\mathrm{reg}}$.
The one related to $\Gamma_{\mathrm{div}}^\mu$ (\ref{eq:Gamma_div})
refers to the ``gauge restoration part":
\begin{align}
M_{\mathrm{nlC}}^{\mathrm{div}} = \frac12 \mathcal{G} \int \frac{d \ell}{2 \pi} \delta^{(4)} (p + \ell k - p' - k') \, \delta (\ell) 
\left[\bar u(p') (- i e) \gamma^\mu {\epsilon_\mu^*}' (k') u(p) \right] = 0 ,
\end{align}  
which vanishes since the $\delta (\ell)$ term, upon $\ell$ integration, enforces the balance equation
$p - p' - k'$ for on-shell momenta, leaving no phase space. The non-zero term, 
related to $\Gamma_{\mathrm{reg}}^\mu$ (\ref{eq:Gamma_reg}), becomes
\begin{align}
M_{\mathrm{nlC}}^{\mathrm{reg}} &= \int \frac{d \ell}{2 \pi} \delta^{(4)} (p + \ell k - p' - k') \nonumber\\
&\times 
\bigg[\bar u(p') (- i e)
\left\{ \left(\Gamma^{\mu \nu}_1 - {\cal P} \frac{\gamma^\mu \alpha_1^\nu}{\ell} \right) B_{1 \nu} (\ell)
+  
\left( \Gamma_2^\mu - {\cal P} \frac{\gamma^\mu \alpha_2}{\ell} \right) B_2 (\ell) \right\}
{\epsilon_\mu^*}' (k') u(p) \bigg] , \\
&= \frac{- i e}{2 \pi k^+}
\delta^{lf} (p - p' - k') \nonumber \\
&\times
\bigg[\bar u(p') 
\bigg\{ \underbrace{\left( \Gamma^{\mu \nu}_1 - {\cal P} \frac{\gamma^\mu \alpha_1^\nu}{\ell_0} \right)}_{\propto e}\ 
\underbrace{ B_{1 \nu} (\ell_0)}_{\propto \frac{a_0}{e} (\cdots)}
+  
\underbrace{\left( \Gamma_2^\mu - {\cal P} \frac{\gamma^\mu \alpha_2}{\ell_0} \right)}_{\propto e^2}\ 
\underbrace{B_2 (\ell_0)}_{\propto \frac{a_0^2}{e^2} (\cdots)} \bigg\}
{\epsilon_\mu^*}' (k') u(p) \bigg] , \label{eq:nlC}
\end{align} 
where in the last two lines the light cone coordinates are employed:
$\delta^{lf} (q) := \frac12 \delta (q^-) \delta^{(2)} (q^\perp)$.
The photon number parameter is $\ell_0 = \frac{(p+k')^2 - m^2}{2 k \cdot p}$. 
The matrix element (\ref{eq:nlC}) is the starting point for many investigations of one-photon nonlinear Compton
in a pulsed plane-wave background. To make this relation explicit we rewrite Eq.~(\ref{eq:G})
by means of (\ref{eq:Amu}) and $a := a_0 m / \vert e \vert$ as
\begin{align}
G(\phi) &= - \mbox{Re} \, \alpha_- \int_{- \infty}^\phi d \phi g(\phi') \exp\{ i (\phi' + \phi_{\mathrm{CEP}}) \} \nonumber \\
           & - \frac12 a^2 \alpha_2 \left( \cos 2 \xi  \int_{- \infty}^\phi d \phi' \, {g(\phi')}^2 \cos (2[\phi' + \phi_{\mathrm{CEP}}])
+ \int_{- \infty}^\phi d \phi' \, {g(\phi')}^2  \right),
\end{align}
where $\alpha_\pm := a \, \alpha_{1 \mu} \, \epsilon_\pm^\mu$ with 
$\epsilon_\pm^\mu \equiv \epsilon_1^\mu \cos \xi \pm i \epsilon_2^\mu \sin \xi$.
Using the abbreviation
\begin{align} \label{eq:Bpm}
B_{1 \pm} (\ell) := \int_{- \infty}^\infty d\phi \, \exp\{ \pm (i [\phi + \phi_{\mathrm{CEP}}]) \} \exp\{- i [\ell \phi + G(\phi)] \}
\end{align}
the phase integrals (cf.\ (\ref{eq:B1}, \ref{eq:B2})) in (\ref{eq:nlC}) can be cast in the form
\begin{align} \label{eq:Bpmeps}
B_1^\mu (\ell) &= \frac 12 a \left\{ \epsilon_+^\mu B_{1 -} (\ell) + \epsilon_-^\mu B_{1 +} (\ell) \right\}, \\
B_2 (\ell)       &= a^2 \int_{-\infty}^\infty d\phi \, g(\phi)^2 \left\{ 1 + \cos 2 \xi \cos  (2[\phi + \phi_{\mathrm{CEP}}]) \right\}
\exp\{ i (\phi + \phi_{\mathrm{CEP}}) \} 
\end{align}
yielding eventually Eq.~(3.26) in \cite{Seipt:2012nad} with many accompanying and subsequent works. 
The one-photon nonlinear Compton process based on the one-vertex diagram
seems to be exhaustively analyzed (cf.\ \cite{Fedotov:2022ely} for a recent review). 
The special setup of multi-color laser background field, e.g.\ the superposition of aligned optical and XFEL beams, 
i.e.\ x-ray scattering at an electron moving in the laser field \cite{Seipt:2013hda,Seipt:2015rda}, 
offer further interesting facets, up to polarization gating to produce 
a mono-energetic $\gamma$ beam \cite{Seipt:2019yds,Valialshchikov:2022ndd}). 
Furthermore, nonlinear Compton has non-perturbative contributions 
(analog to nonlinear Breit-Wheeler), and \cite{HernandezAcosta:2020agu}
shows how to isolate them. 

With respect to power counting of $e$ and $a_0$, the assignments are displayed in Eq.~(\ref{eq:nlC}),
where $(\cdots )$ stands for the series expansion of the $\exp\{ i G\}$ term.

\subsection{Nonlinear Breit-Wheeler}\label{sect:nlBW}

The matrix elements of nonlinear Breit-Wheeler (nlBW) refer to
\begin{align}
\underbrace{\includegraphics[width=0.22\textwidth,valign=c]{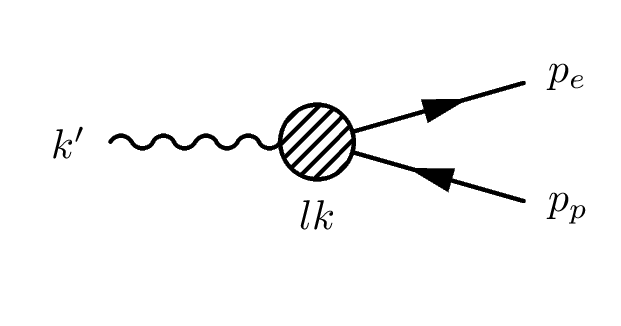}}_{M_{\mathrm{nlBW}}} 
=
\underbrace{\includegraphics[width=0.22\textwidth,valign=c]{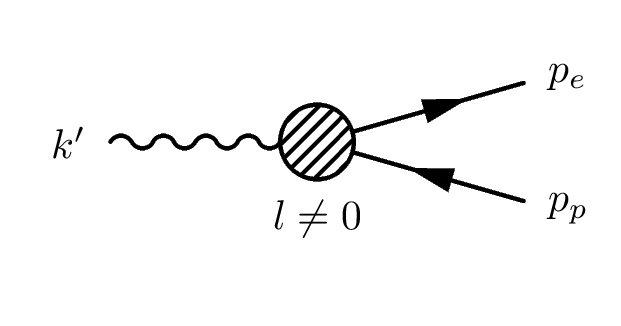}}_{M_{\mathrm{nlBW}}^{\mathrm{reg}}} 
+  
\underbrace{\includegraphics[width=0.22\textwidth,valign=c]{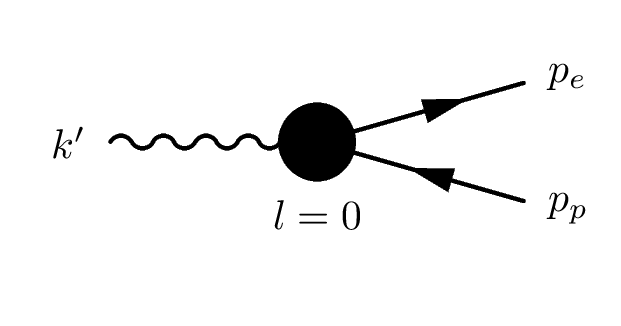}}_{M_{\mathrm{nlBW}}^{\mathrm{div}}} 
\end{align}
and read, analog to the nonlinear Compton as crossing channel Eqs.~( \ref{eq:nlC_0} - \ref{eq:nlC}),
\begin{align}
M_{\mathrm{nlBW}} &= \int \frac{d \ell}{2 \pi} 
[ \bar u(p_e) (-ie) \Gamma^\mu (\ell , - p_p, p_e)  {\epsilon_\mu}'  \nu(p_p) ]
\delta^{(4)} (k' + \ell k - p_p - p_e)  \\
&= \frac{- i e}{2 \pi k^+} [\bar u(p_e) \Gamma^\mu (\ell_0, -p_p, p_e) {\epsilon_\mu}' \nu(p_p)] 
\bigg\vert_{\ell_0 = \frac{(p_e + p_p)^2}{2 k \cdot k'}} ,  \\
M_{\mathrm{nlBW}}^{\mathrm{div}} &= \frac{- ie}{2 \pi k^+} \pi \mathcal{G}(-p_p, p_e)
\delta(\ell_0) \, \delta^{lf} (k' - p_p -p_e) [\bar u(p_e) \gamma^\mu  {\epsilon_\mu}'  \nu(p_p) ] \to 0 , \\
M_{\mathrm{nlBW}}^{\mathrm{reg}} &= \frac{-ie}{2 \pi k^+} \delta^{lf} (k' - p_p -p_e)\nonumber\\
& \phantom{==}\times\bigg[ \bar u(p_e) \bigg\{ \left( \Gamma_1^{\mu \nu} (-p_p, p_e) - \mathcal{P} \frac{\gamma^\mu \alpha_1^\nu(-p_p,p_e)}{\ell} \right)
B_{1 \, \nu} (\ell , - p_p, p_e)\nonumber  \\
& \hspace*{2.8cm} + \left.
\left( \Gamma_2^\mu (-p_p,p_e) - \mathcal{P} \frac{\gamma^\mu \alpha_2(-p_p,p_e)}{\ell} \right)
B_2(\ell ,  - p_p, p_e) \right\}  
{\epsilon_\mu}' \nu (p_p) \bigg],
\end{align}
where $M_{\mathrm{nlBW}}^{\mathrm{div}} \to 0$ is a result of combining the $\delta$ distributions
$\delta(\ell_0) \, \delta^{lf} (k' - p_p -p_e) \to \delta^{(4)} (k' - p_e -p_e)$ which can not be satisfied
by on-shell momenta.

While, in nonlinear Compton, the initial electron momentum may be zero, $\vec p = 0$, 
due to the action of the external classical field, the electron can emit a real photon(s),
whatever the external-field central-frequency $\omega > 0$ is.
One can imagine this as shaking off photons due to the quiver motion in the external field.
The crossing channel, i.e.\ nonlinear Breit-Wheeler as one-photon decay,
$\gamma' \to e_L^+ e_L^-$ with matrix element $\propto {k_\mu}' [v_{p'} \Gamma^\mu \bar u_p]$),
faces a severe threshold, making nonlinear Compton and
nonlinear Breit-Wheeler quite distinctive, even the amplitudes are related by
crossing symmetry.
The balance equations in the monochromatic case read
\begin{equation}
\ell k \pm k' = \pm q + q'
\end{equation}
with quasi-momenta $q = p + a_0^2 m^2 /(2 p \cdot k)$
and $q' = p' + a_0^2 m^2 /(2 p' \cdot k)$ which facilitate
$q^2 = {q'}^2 = m_*^2$, lead to 
\begin{eqnarray}
\ell k \cdot k' &=& q \cdot q + m_*^2 \quad \mbox{(nlBW,\, upper \, sign),}\\
\ell k \cdot p &=& k' \cdot q \quad \mbox{(nlC, \, lower \, sign)} .
\end{eqnarray} 
Explication for nonlinear Compton (head-on laser-electron collisions) reads
\begin{eqnarray}
k &=& (\omega, \omega, 0, 0), \\
p &=& (m \cosh y, - m \sinh y, 0, 0),\\
k' &=& (\omega',\omega' \cos \Theta', \omega' \sin \Theta',0),\\
p' &=& (E',\sqrt{{E'}^2 -m^2 - {\omega'}^2 \sin^2 \Theta'}, - \omega' \sin \Theta',0),
\end{eqnarray}
where $E = m \cosh y$ and $\vert \vec{p}\, \vert = m \sinh y$ relates energy $E$ and 
momentum $\vec p$, $E^2 + {\vec p}^{\,2} = m^2$,
with rapidity $y$, and
\begin{eqnarray}
\omega' (\ell, \cos \Theta') &=& \frac{\ell \omega}{1 + e^{-y} \kappa (1 -\cos \Theta')},
\quad \kappa := \ell \frac{\omega}{m} -\sinh y + \frac12 a_0^2 e^{-y}, \\
E' &=& \frac{m^2 - \omega' (m- \omega')(1 -\cos \Theta')}{m - \omega' (1 - \cos \Theta')} 
\end{eqnarray}
express $\omega'$ and $E'$ as a function of $\cos \Theta'$. 
Forward (backward) scattering is defined by $\cos \Theta' = 1$ ($-1$).
The $out$-electron angle is determined by $\sin \theta' = - \omega' \sin \Theta'/ \vert p' \vert$. 

In nonlinear Breit-Wheeler, the quasi-momenta $q$ and $q'$ symmetrically enter the corresponding
kinematic equations. The threshold energy, for the monochromatic case, 
is determined by $k \cdot k' = 2 m_*^2$. However, the
sub-threshold pair production is enabled in short pulses, as emphasized in
\cite{Titov:2020taw,Titov:2019kdk,Titov:2013kya,Nousch:2012xe,Titov:2012rd}.
Temporal double pulses or bichromatic pulses enhance further the pair rate,
as suggested in \cite{Titov:2018bgy,Otto:2016fdo,Nousch:2015pja}.

While Compton has a classical analog (shaking off the e.m.\ field accompanying
an accelerated charge in the form of asymptotically outgoing waves), 
Breit-Wheeler is said to be a quantum process, i.e.\ ``converting light into matter".
A particularly interesting aspect is the relation to vacuum birefringence,
see \cite{Borysov:2022cwc}, which is experimentally searched for
in dedicated and highly specialized \& optimized set-ups,
e.g.\ pursued by HIBEF 
\cite{Ahmadiniaz:2022nrv,Ahmadiniaz:2020lbg,Schlenvoigt:2016jrd,Heinzl:2006xc}. 

\section{Two-vertex processes/four-point amplitude}\label{sect:two_vertex}

The two two-vertex diagrams have the symbolic matrix elements
(i) ${\cal M} \sim [ \bar u_L(p') \Gamma^\mu S_F(Q) \Gamma^\nu u_L(p) ] E_\mu(k_1) E_\nu(k_2)$ and
(ii) ${\cal M} \sim [\bar u_L(p') \Gamma^\mu u_L(p)] {\cal D}_{\mu \nu} [\bar u_L(P') \Gamma^\nu u_L(P)]$
with $S_F$ and ${\cal D}_{\mu \nu}$ as Fermion and photon propagators.
They have one (i) and two (ii) tied Fermion lines.
Again, depending on the orientation of the four-momenta, several processes 
related by crossing symmetry are conceivable:\\
(i) $e_L^- \to {e_L^-}' + \gamma_1 + \gamma_1$ (nonlinear two-photon Compton, 
cf.\ \cite{Lotstedt:2009zz,Loetstedt:2009zz,Seipt:2012tn,Mackenroth:2012rb}),
$e_L^- + \gamma \to {e_L^-}' + \gamma'$ (nonlinear Compton scattering, i.e.\
x-ray Compton scattering at an electron moving in a non-aligned optical laser),
$\gamma_1 + \gamma_2 \to e_L^- + e_L^+$ (nonlinear two-photon Breit-Wheeler)
and time-reversed processes as well, in particular $e_L^- + e_L^+ \to \gamma_1 + \gamma_2$; \\ 
(ii) $e_{1L}^- + e_{2L}^- \to {e_{1L}^-}' + {e_{2L}^-}'$ (nonlinear M{\o}ller scattering),
$e_{L}^- \to  {e_L^-}' + {e_L^-}'' + {e_L^+}$ (nonlinear trident)
and several crossing channels as well
(e.g.\ nonlinear Bhabha scattering with $s$ and $t$ channel diagrams).
Also, the involvement of two different lepton species is conceivable, e.g.\ electrons and muons.

We explicate now our momentum space Furry-picture Feynman-rules for
nonlinear two-photon Compton (subsection \ref{sect:nl2C})
and nonlinear M{\o}ller (subsection \ref{sect:nlM}).
 
\subsection{Two-photon nonlinear Compton}\label{sect:nl2C}

\subsubsection{Diagrams and matrix element}

The two-photon nonlinear Compton (2nlC)
$e^- (p) + \text{laser} \to \gamma(k'_1) + \gamma(k'_2) + e^-(p') + \text{laser}$
as a two-vertex tree-level diagram 
\begin{align}\label{eq:Feyndiags_general}
\includegraphics[width=0.22\textwidth,valign=c]{sf_2gamma_compton1} \qquad + \quad \includegraphics[width=0.22\textwidth,valign=c]{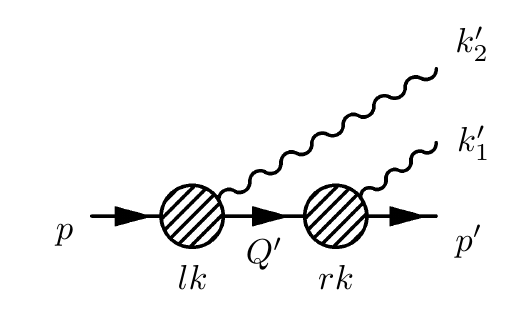},
\end{align}
requires a somewhat more intricate treatment, see
\cite{Lotstedt:2009zz,Loetstedt:2009zz,Seipt:2012nad,Seipt:2012tn,Mackenroth:2012rb},
despite the simple matrix element (direct term, i.e.\ left diagram; the exchange term, i.e.\ right diagram,
is to be processed analogously)
\begin{align} \label{nl2C}
S_{\mathrm{2nlC}} &= \int \frac{d \ell}{2 \pi} \frac{d r}{2 \pi} \frac{d^{(4) Q}}{(2 \pi)^4}
\delta^{(4)} (p + \ell k - k_1' - Q) 
\delta^{(4)} (Q + r k - k_2' - p') \nonumber \\
&\times
\left[ \bar u(p') 
(-ie) \Gamma^\mu (r, Q, p' \vert k)  {\epsilon_\mu^*}' (k_2')
S_F(Q)   
(-ie) \Gamma^\nu (\ell, p, Q \vert k)  {\epsilon_\nu^*}' (k_1') u(p) \right] ,
\end{align} 
where $S_F = \frac{i(\slashed Q + m)}{Q^2 - m^2 + i \epsilon}$ is Feynman's free Fermion propagator.
Collinear divergence and IR behavior as well as the on/off-shell behavior of the Fermion propagator provide some challenges.
In addition, the soft-photon theorem (cf.\ \cite{soft_photon_ALICE3} for contemporary reasoning)
might be explicated here.

We consider now only the structure of the direct matrix element (\ref{nl2C}), where $p$ and $p'$ denote the momenta
of the in- and out-going electrons, and $k_{1, 2}'$ are momenta of the outgoing photons
with polarization four-vectors ${\epsilon_{\mu, \nu}'}^* (k_{1, 2}')$;
$Q$ refers to the intermediate electron. 
Using one of the \( \delta \)-distributions, 
the integral over the intermediate-electron momentum \( Q \) can be solved analytically:
\begin{align}\label{eq:matrix_element_general}
S_{\mathrm{2nlC}}&= \frac{-e^2}{(2 \pi)^6}\int \mathrm{d} \ell \int \mathrm{d} r\,  
\delta^{(4)}(p + (r+\ell)k - k_1' - k_2'-p')  \varepsilon^{\prime *}_{2 \mu}(k_2') \varepsilon^{\prime *}_{1  \nu}(k_1')\nonumber \\
& \times [\overline u(p') \Gamma ^ \mu(r, Q, p'\vert k) \, S(Q)\,\Gamma ^ \nu(\ell, p,Q \vert k)  u(p) ],
\end{align}
where \( Q \) is now related to the external momenta and the photon number parameters via
\begin{align}\label{eq:fixed_intermediate_momentum}
p+ \ell k - k_1' = Q = k_2' + p' - rk.
\end{align}
The remaining \( \delta \)-distribution in \eqref{eq:matrix_element_general} can be used to solve one of the photon-number parameter integrals by applying light-cone coordinates to the involved momenta:
\begin{align}\label{eq:delta_in_lcc}
\delta^{(4)}(p+ (r+\ell)k - k_1' - k_2' - p') = \delta^{\mathrm{lf}}(p - k_1' - k_2' - p') \,
\delta(p^++ (r+\ell)k^+ - k_1^{\prime +} - k_2^{\prime +} - p^{\prime +}), 
\end{align}
where \( \delta^{\mathrm{lf}(q)} = \frac{1}{2}\delta^{(2)}(q^\perp) \, \delta(q^-) \). 
The second \( \delta \)-distribution in \eqref{eq:delta_in_lcc} can be used to solve one of the integrals over the photon number parameter, e.g.\ the \( r \)-integral, which leads to
\begin{align}\label{eq:matrix_element_final_general}
S_{\mathrm{2nlC}} &= \frac{-e^2}{(2 \pi)^6 k^+}  \delta^{\mathrm{lf}}(p - k_1' - k_2' - p')   
\varepsilon^{\prime *}_{2 \mu}(k_2')\varepsilon^{\prime *}_{1 \nu}(k_1') \nonumber \\
&\times  \int \mathrm{d} \ell \, 
[ \overline u(p') \Gamma ^ \mu(r_\ell, Q, p'\vert k) \, S(Q)\,\Gamma ^ \nu(\ell, p,Q \vert k)  u(p) ],
\end{align}
where the two photon number parameters are no longer independent, but related by \( r_\ell := \ell_0 - \ell \) with
\begin{align}\label{eq:pNum0}
\ell_0 := \frac{k_1^{\prime +} + k_2^{\prime +} + p^{\prime +} - p^+}{k^+} = 
\frac{(k_1' + k_2' + p')^2 - m^2}{2p \cdot k}.
\end{align}

\subsubsection{Singularity structures}\label{sec:singularity_structures}

One obvious source of singularities is the vanishing denominator of the field-free electron propagator  
\( S_F(Q) \). Therefore, the general resonance condition is given if the intermediate electron goes on-shell, i.e.
$Q^2 = m^2$.
One convenient way to keep control of the propagator singularity is to define the virtuality of the intermediate electron: 
\( v:= Q^2 - m^2 \). Deploying Eq. \eqref{eq:fixed_intermediate_momentum}, 
the virtuality is a function of either one of the two photon-number parameters:\footnote{There is only one \emph{virtuality} for the intermediate electron, \( \nu:= Q^2 - m^2 \). However, here we phrase the dependence of the virtuality on the photon-number parameters as if they are independent, because the choice, which of the photon-number parameter integral in equation \eqref{eq:matrix_element_general} one wants to solve, is arbitrary. Consequently, both of the \emph{definitions} \eqref{eq:virtuality_l} and \eqref{eq:virtuality_r} are equivalent, if and only if \( r \equiv r_\ell = \ell_0 - \ell \).}
\begin{align}
\nu_\ell &:= Q^2(\ell) - m^2 = (p-k_1')^2 + 2 \ell k \cdot (p-k_1') , \label{eq:virtuality_l}  \\
\nu_r &:= Q^2(r) - m^2 = (k_2' + p')^2 - 2r k \cdot (k_2' + p'), \label{eq:virtuality_r}
\end{align}
where we have \( \nu_\ell \equiv \nu_{r_\ell = \ell_0 - \ell} \). 
The resonance condition 
\( \nu:= Q^2 - m^2 \) is then equivalent to \( \nu_\ell = \nu_r = 0 \). 
Written with the virtuality \( \nu_\ell \), the denominator of the field-free fermion propagator reads
\begin{align}\label{eq:elec_prop_l}
\frac{1}{Q^2 - m^2 + i \epsilon} = \frac{1}{ \nu_\ell + i \epsilon} = \frac{1}{2 k \cdot (p-k_1)} 
\left(\frac{1}{\ell_{\mathrm{on}}- \ell + i \epsilon}\right),
\end{align}
where we employ the replacement \( \frac{ \epsilon}{2 k \cdot (p-k_1')} \to \epsilon  \) and we use the abbreviation 
\begin{align}\label{eq:pNum_onshell}
 \ell_{\mathrm{on}} = \frac{(p-k_1')^2 - m^2}{2k \cdot (p-k_1')} \neq 0. 
\end{align}
Analogously, written with the virtuality \( \nu_r \), one gets
\begin{align}
\frac{1}{Q^2 - m^2 + i \epsilon} = \frac{1}{ \nu_r + i \epsilon} = \frac{1}{2 k \cdot (k_2' + p')} 
\left(\frac{1}{r_{\mathrm{on} }- r + i \epsilon}\right)
\end{align}
with \( r_{\mathrm{on}} = \frac{(k_2' + p')^2 - m^2}{2 k \cdot (k_2' + p')} \neq 0 \). 
Consequently, the singularity structure of the electron propagator in the matrix element \eqref{eq:matrix_element_final_general} is directly related to the values of the photon number parameters at the respective vertex, 
i.e.\ \( \nu_\ell = 0 \Leftrightarrow \ell = \ell_{\mathrm{on}}\) or equivalently 
\(  \nu_r = 0 \Leftrightarrow r =r_{\mathrm{on}}\). As we will show in the sequel, 
these types of singularities are the only ones, which may appear in the matrix element \eqref{eq:matrix_element_final_general}.

\subsubsection{Asymptotically vanishing field case}

As illustrated in Section  \ref{dressed_vertex_decomposition}
in the case of plane-wave pulses, i.e.\ asymptotically vanishing fields, 
the manifestly gauge-invariant dressed vertex function $\Gamma^\mu$
decomposes into a finite $\Gamma_{\mathrm{reg}}^\mu$ and a gauge-restoration $\Gamma_{\mathrm{div}}^\mu$
part. Consequently, inserting the decomposition 
by Eqs.~(\ref{eq:Gamma_decomp} - \ref{eq:Gamma_reg})
into Eq.~\eqref{eq:matrix_element_final_general}, the matrix element of strong-field two-photon-Compton scattering 
becomes
\begin{align}
\underbrace{\includegraphics[width=0.17\textwidth,valign=c]{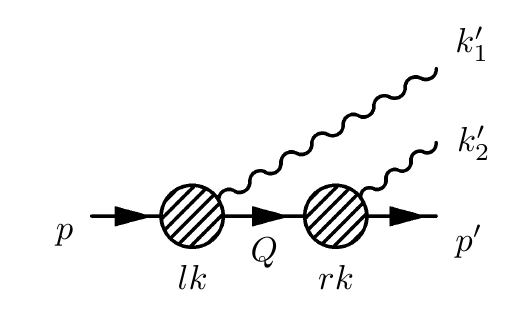}}_{=S_{\mathrm{2nlC}}}=
\underbrace{\includegraphics[width=0.17\textwidth,valign=c]{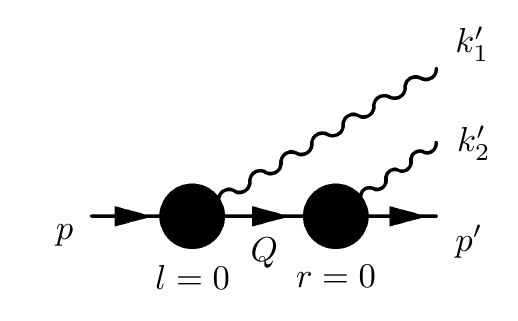}}_{=:S_{\mathrm{2nlC}}^{(0)}}+
\underbrace{\includegraphics[width=0.17\textwidth,valign=c]{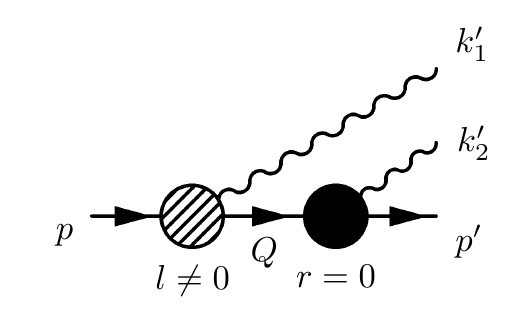}}_{=:S_{\mathrm{2nlC}}^{(11)}}+
\underbrace{\includegraphics[width=0.17\textwidth,valign=c]{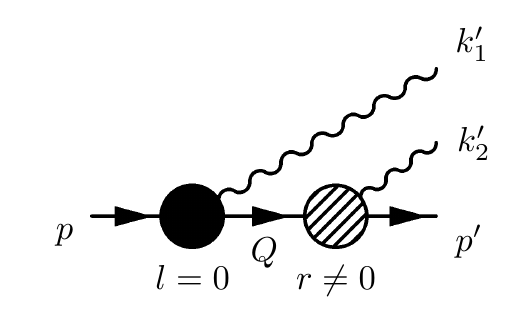}}_{=:S_{\mathrm{2nlC}}^{(12)}} +
\underbrace{\includegraphics[width=0.17\textwidth,valign=c]{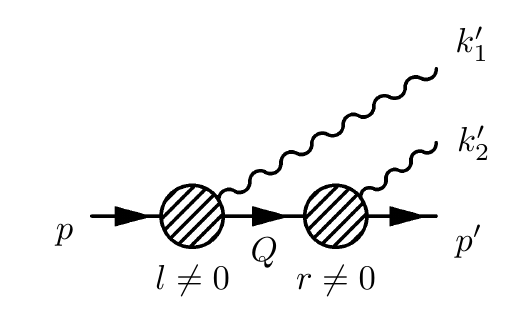}}_{=:S_{\mathrm{2nlC}}^{(2)}},\label{eq:sf_2gammaC_decomp_diag}
\end{align}
where the vertex structure of the particular matrix element introduces constraints on the respective photon-number parameter. 
We consider term by term:\\
(i) \emph{No energy-momentum transfer} --
    The first part of \eqref{eq:sf_2gammaC_decomp_diag} corresponds to the case, where both vertices are given as the divergent part of the dressed vertex, which implies that for this diagram, there is no energy-momentum transfer w.r.t. the background field on neither of the vertices. This follows also directly from the corresponding part of the 2nlC matrix element, which reads
\begin{align}
S_{\mathrm{2nlC}}^{(0)}&= \frac{-e^2 \pi^2}{(2 \pi)^6 k^+}  
\delta^{\mathrm{lf}}(p - k_1' - k_2' - p')   \varepsilon^{\prime *}_{2 \mu}(k_2')\varepsilon^{\prime *}_{1 \nu}(k_1')\nonumber\\ 
        &\phantom{==}\times \int \mathrm{d} \ell \, 
[\overline u(p') \delta(r_\ell) \mathcal{G}(Q, p')  \gamma^\mu  S(Q)  \gamma^\nu \delta(\ell) \mathcal{G}(p,Q)  u(p) ] \\
        &=\frac{-e^2 \pi^2}{(2 \pi)^6 k^+}  \delta^{\mathrm{lf}}(p - k_1' - k_2' - p')\delta(r_0)\nonumber \\
        &\phantom{==}\times \mathcal{G}(Q, p') \mathcal{G}(p,Q)   
\varepsilon^{\prime *}_{2 \mu}(k_2')\varepsilon^{\prime *}_{1 \nu}(k_1') 
\left[\overline u(p')   \gamma^\mu  S(Q)  \gamma^\nu   u(p)\right] , \label{eq:matEl_s0}
        \end{align}
    where the \( \delta \)-distributions from the gauge-restoration parts are solved by \( \ell = 0 = r_0 \equiv \ell_0 \). 
The remaining \( \delta \)-distributions only depend on the external particles. Therefore, using \eqref{eq:delta_in_lcc} and \eqref{eq:pNum0} leads to
        \begin{align}
        \frac{1}{k^+}\delta^{\mathrm{lf}}(p - k_1' - k_2' - p') \, \delta(\ell_0) = \delta^{(4)}(p - k_1' - k_2' - p').
        \end{align}
    However, there is no physical phase space solving \( p - k_1' - k_2' - p'=0 \) with all on-shell momenta. Furthermore, for the  first part of \eqref{eq:sf_2gammaC_decomp_diag}, the virtualities read
    \begin{align}
    \nu_{\ell=0} &= (p-k_1')^2 - m^2 \neq 0,\\
    \nu_{r=0} &= (k_2' + p')^2 - m^2 \neq 0.
    \end{align}
Therefore, there is no singularity to \emph{cancel} the vanishing phase space, 
thus there is no contribution of \( S_{\mathrm{2nlC}}^{(0)} \) to the matrix element.

(ii) \emph{Contribution from the left vertex} --
    The second term in the decomposition \eqref{eq:sf_2gammaC_decomp_diag} represents the case, where at the left vertex, the photon-number parameter does not vanish, i.e. \( \ell \neq 0 \), whereas, on the right vertex, the photon-number parameter is identically zero: \( r_\ell = \ell_0 - \ell=0 \). 
The corresponding part of the strong-field 2nlC matrix element is given by
\begin{align}\label{eq:matEl_S11}
S_{\mathrm{2nlC}}^{(11)}&= \frac{-e^2 \pi}{(2 \pi)^6 k^+}  
\delta^{\mathrm{lf}}(p - k_1' - k_2' - p')   \varepsilon^{\prime *}_{2 \mu}(k_2')\varepsilon^{\prime *}_{1 \nu}(k_1')\nonumber\\
        &\phantom{==}\times\int \mathrm{d} \ell \, [\overline u(p') \delta(r_\ell) \mathcal{G}( Q, p') \gamma ^\mu 
S(Q) \Gamma ^ \nu_{\mathrm{reg}}(\ell, p,Q \vert k)  u(p) ]\\
        &= \frac{-e^2 \pi}{(2 \pi)^6 k^+}  \delta^{\mathrm{lf}}(p - k_1' - k_2' - p')   \varepsilon^{\prime *}_{2 \mu}(k_2')\varepsilon^{\prime *}_{1 \nu}(k_1')\nonumber\\
        &\phantom{==}\times \mathcal{G}( Q, p') \left[\overline u(p')  \gamma ^\mu 
S(Q) \Gamma^ \nu_{\mathrm{reg}}(\ell_0, p,Q \vert k)  u(p)\right],
\end{align}
 where \( \ell_0 \) is given by Eq.~\eqref{eq:pNum0} and \( \Gamma ^{\mu}_{ \mathrm{reg}}\) is the finite part of the dressed vertex defined in Eq.~\eqref{eq:Gamma_reg}. 
The Cauchy principal value operator in \( \Gamma ^{\mu}_{ \mathrm{reg}}\) ensures \( l_0 \neq 0 \). 
For this part of the matrix element, the virtualities \eqref{eq:virtuality_l} and \eqref{eq:virtuality_r} read
\begin{align}
\nu_{r_\ell=0} &= (k_2' + p')^2 - m^2 \neq 0,\\
\nu_{\ell=\ell_0} &\equiv \nu_{r=0} \neq 0,
\end{align}
thus, there is no singularity in \(  S_{\mathrm{2nlC}}^{(11)} \).

(iii) \emph{Contribution from the right vertex} -- 
    The third term in the decomposition \eqref{eq:sf_2gammaC_decomp_diag} represents the case, where at the right vertex, the photon-number parameter does not vanish, i.e. \( r_l\neq 0 \), whereas, at the left vertex, the photon-number parameter is identically zero: \( \ell=0 \). The corresponding part of the strong-field 2nlC matrix element is given as 
\begin{align}
        S_{\mathrm{2nlC}} ^{(12)} &= \frac{-e^2 \pi}{(2 \pi)^6 k^+}  \delta^{\mathrm{lf}}(p - k_1' - k_2' - p')   \varepsilon^{\prime *}_{2 \mu}(k_2')\varepsilon^{\prime *}_{1 \nu}(k_1')\nonumber\\
        &\phantom{==} \times \int \mathrm{d} \ell \, [\overline u(p') \Gamma ^ \mu_{ \mathrm{reg}}(r_\ell, Q, p'\vert k) \, S(Q)\, \gamma _{\mu} \delta( \ell) \mathcal{G}(p,Q )  u(p) ]\\
        &=  \frac{-e^2 \pi}{(2 \pi)^6 k^+}  \delta^{\mathrm{lf}}(p - k_1' - k_2' - p')   \varepsilon^{\prime *}_{2 \mu}(k_2')\varepsilon^{\prime *}_{1 \nu}(k_1')\nonumber\\
        &\phantom{==} \times  \mathcal{G}(p,Q ) \left[ \overline u(p') \Gamma ^ \mu_{ \mathrm{reg}}(r_0, Q, p'\vert k) \, S(Q)\, \gamma _{\mu}   u(p)\right],
\end{align}
where \( r_0 \equiv \ell_0 \) is given by Eq.~\eqref{eq:pNum0} and  \( \Gamma ^{\mu}_{ \mathrm{reg}}\) 
is again the regularized finite part of the dressed vertex defined in 
Eq.~(\ref{eq:Gamma_reg})
and the Cauchy principal value operator in 
\( \Gamma ^{\mu}_{ \mathrm{reg}}\) ensures \( r_0 \neq 0 \). 
For this part of the matrix element, the virtualities \eqref{eq:virtuality_l} and \eqref{eq:virtuality_r} read
\begin{align}
\nu_{\ell=0} &= (p-k_1')^2 - m^2 \neq 0,\\
\nu_{r=r_0 = \ell_0} &\equiv \nu_{\ell=0} \neq 0,
\end{align}
thus, there is no singularity in \(  S_{\mathrm{2nlC}}^{(12)} \).

(iv) \emph{On-shell- and off-shell contributions} -- 
    The fourth term in the decomposition \eqref{eq:sf_2gammaC_decomp_diag} represents the case, where on both vertices the photon-number parameters are non-zero, i.e. \( \ell \neq 0\) and \(r_\ell \neq 0\). 
The corresponding part of the strong-field 2nlC matrix element is given as 
\begin{align}
S_{\mathrm{2nlC}}^{(2)}&= \frac{-e^2}{(2 \pi)^6 k^+}  
\delta^{\mathrm{lf}}(p - k_1' - k_2' - p')   \varepsilon^{\prime *}_{2 \mu}(k_2')\varepsilon^{\prime *}_{1 \nu}(k_1')\nonumber \\
        &\phantom{==}\times\int \mathrm{d} \ell \, [\overline u(p') \Gamma ^ \mu_{ \mathrm{reg}}(r_\ell, Q, p'\vert k) \, 
S(Q)\,\Gamma ^ \nu_{ \mathrm{reg}}(\ell, p,Q \vert k)  u(p) ] ,
\end{align}
where \( \Gamma ^{\mu}_{ \mathrm{reg}}\) is the finite part of the dressed vertex defined in 
Eq.~(\ref{eq:Gamma_reg}).
The Cauchy principal value operator in \( \Gamma ^{\mu}_{ \mathrm{reg}}\) 
ensures \(\ell\neq 0 \neq r_\ell\). 
However, according to Eq.~\eqref{eq:elec_prop_l}, the condition \( \ell = \ell_{ \mathrm{on}} \), 
with \(  \ell_{ \mathrm{on}} \) defined in \eqref{eq:pNum_onshell}, is equivalent to \(Q^2 - m^2= 0\) 
inducing the propagator \(S_F(Q) = \frac{i(\slashed Q + m)}{Q^2 - m^2 + i \epsilon}\) 
to diverge in the limit \( \epsilon \to 0\). 
This divergence can be handled by applying the Sokhotski–Plemelj theorem:
\begin{align}\label{eq:sokhotski}
\lim_{\epsilon\to 0^+} \frac{1}{Q^2 - m^2 + i \epsilon} = -i \pi \delta(Q^2 - m^2) 
+ \mathcal{P} \frac{1}{Q^2 - m^2},
\end{align}
where, on the r.h.s.,  in the first term the denominator of the propagator is removed and \(Q\) is set on-shell, 
i.e.\ \(Q^2 \overset{!}{=} m^2\), or equivalently \( \ell \overset{!}{=} \ell_{\mathrm{on}}\), 
where \( \ell_{\mathrm{on}} \) is given in Eq.~\eqref{eq:pNum_onshell}. 
In the second term of the r.h.s.\ of equation \eqref{eq:sokhotski}, the Cauchy principal value operator ensures that the denominator of the propagator never vanishes, which removes the singularity caused by \(Q^2 = m^2\). 
In the language of virtualities introduced in Section \ref{sec:singularity_structures}, 
this means for the first term of the r.h.s.\ of \eqref{eq:sokhotski}, the virtualities read \( \nu_\ell = \nu_{r_\ell} =0\ \), 
i.e.\ the intermediate electron goes on-shell. However, since for this term, the denominator of the propagator is removed, 
the on-shell \emph{intermediate} electron does not induce a pole. 
The second term of the r.h.s.\ of Eq.~\eqref{eq:sokhotski} does not induce a pole neither,
because the Cauchy principal value operator protects the denominator of the electron propagator from vanishing, 
i.e.\ \( \mathcal{P}\frac{1}{Q^2 - m^2} = \mathcal{P} \frac{1}{ \nu} \), 
which excludes the value \( \nu = 0 \) from the  integration region. 
Therefore,  there is no singularity in \(  S_{\mathrm{2nlC}}^{(2)} \).

In summary, it can be said, therefore, that in the case of asymptotically vanishing background fields, \( \Delta \phi < \infty \), the matrix element \eqref{eq:matrix_element_final_general} of two-photon Compton scattering has no singularities 
except for a single light-front \( \delta \)-distribution, which ensures the conservation of the transverse 
and the minus components of the external momenta.

\subsubsection{Infinitely extended plane wave/Oleinik resonances} \label{sec:Oleinik}

The special case of a monochromatic plane-wave background field that is infinitely extended 
(see Eq.~(\ref{eq:Amu}) with \( \Delta \phi\to \infty \) or $g = 1$) should be considered separately. 
Now, the two-photon Compton matrix element \eqref{eq:matrix_element_final_general} reads
\begin{align}\label{eq:matrix_element_IPW}
S_{\mathrm{2nlC}}^{\mathrm{IPW}}&= \frac{-e^2}{(2 \pi)^6 k^+}  
\delta^{\mathrm{lf}}(p - k_1' - k_2' - p')   \varepsilon^{\prime *}_{2 \mu}(k_2')\varepsilon^{\prime *}_{1 \nu}(k_1')\nonumber  \\
     &\phantom{==} \times \int \mathrm{d} \ell \, \sum_{n,n'=-\infty}^\infty \delta\left( \ell - \beta(p,Q)- n\right) 
\delta\left( r_l - \beta(p', Q)- n'\right)\nonumber \\
     &\phantom{=======}\times [\overline u(p') \Gamma ^ \mu_{\mathrm{IPW,n'}}(r_\ell, Q, p'\vert k) \, 
S(Q)\,\Gamma ^ \nu_{\mathrm{IPW,n}}(\ell, p,Q \vert k)  u(p) ],
\end{align}
where the mode-wise dressed vertex function \( \Gamma^\mu_{\mathrm{IPW,n}} \) is given in 
Eq.~(\ref{eq:Gamma_IPW}).
We mention that the functions  \( \beta(p,Q) \) and \( \beta(Q,p') \) do not depend on the photon-number parameter \( \ell \), 
since \( \beta(p,Q) = \beta(p,p - k_1') \) and \(  \beta(Q,p') =  \beta(p' + k_2',p') \), respectively. 
Therefore, exchanging the integration and the two summations, one can solve the integral by using one of the \( \delta \)-distributions, e.g.\ the first one, which leads to \( \ell = \ell^{ \mathrm{reso}}_{n} = \beta(p,p-k_1') + n \). 
Then, the respective virtuality \eqref{eq:virtuality_l} results in
\begin{align}
\nu_{\ell=\ell^{ \mathrm{reso}}_{n}} = -k_1' \cdot (p + \{\beta_p + n\} k) + n k \cdot p,
\end{align}
where we use the abbreviation \( \beta_p =  \frac{a^2 e^2}{2} \frac{1}{k\cdot p}  \). 
Finally, if one employs the resonance condition \(  \nu_{\ell=\ell^{ \mathrm{reso}}_{n}} = 0 \), 
for the emitted photon with four-momentum \( k_1^{\prime \mu} = \omega_1' n_1^{\prime \mu} \), one finds for the resonance energy\footnote{This formula for the resonance energy is already known, see, e.g., 
\cite{Loetstedt:2009zz,Seipt:2013hda}. }\begin{align}
\omega_{1,n}^{\prime \mathrm{reso}} = \frac{n k \cdot p}{(p + \{\beta_p + n) k \} \cdot n_1'}.
\end{align}
Singularities of this type are called Oleinik resonances 
\cite{Oleinik:1967,Oleinik:1968}, 
which were already identified for the nonlinear two-photon Compton process in
\cite{Seipt:2013hda,Lotstedt:2009zz,Loetstedt:2009zz}.
For further investigations of the diagrams (\ref{eq:Feyndiags_general}) w.r.t.\ Oleinik resonances 
we refer the interested reader to
\cite{Roshchupkin:2022wpw,Roshchupkin:2022hfq,Roshchupkin:2021vut,Dubov:2020kwi}
and further citations therein,
where one of the photon lines is attributed to a ``field photon" of a nucleus. 

The multi-photon nonlinear Compton with more than two vertices,
e.g.\ \cite{Lotstedt:2012zz,Lotstedt:2013uya},
perpetuates this line of arguments and offers a test bed of gluing techniques, 
such as developed in \cite{Dinu:2018efz,Dinu:2019pau}.

\subsection{Nonlinear M{\o}ller}\label{sect:nlM}

As a further application of the momentum-space Furry-picture Feynman rules 
to two-vertex processes we consider
nonlinear M{\o}ller scattering (nlM), i.e.\ $e_L^-(p_1) + e_L^-(p_2) \to e_L^-(p_1') + e_L^-(p_2')$ in the 
laser background field (\ref{eq:Amu}). 
Here, the Oleinik resonances are attributed to the on-shell contributions of the photon propagator, cf.\ \cite{Bos:1978th}.
The leading-order tree level two-vertex diagrams are
\begin{align}\label{eq:Feyndiags_general}
\underbrace{\includegraphics[width=0.22\textwidth, valign=c]{sf_moller1}}_{S_{\mathrm{nlM}}^{\mathrm{d}}}
\hspace{0.5cm} -\hspace{0.5cm} 
\underbrace{\includegraphics[width=0.22\textwidth, valign=c]{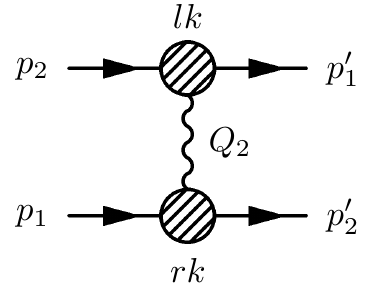}}_{S_{\mathrm{nlM}}^{\mathrm{ex}}} .
\end{align}
The direct term (first diagram) corresponds to the matrix element
\begin{align}
S_{\mathrm{nlM}} &= \int \frac{d \ell}{2 \pi} \frac{d r}{2 \pi} \frac{d^4 Q}{(2 \pi)^4}
\left[ \bar u(p_1') (-ie) \Gamma^\mu (r, p_1, p_1' \vert k) u(p_1) \right]
\delta^{(4)} (p_1 + r k - p_1' - Q) \nonumber\\
&\times \hspace*{1.3cm}
{\cal D}_{\mu \nu} (Q)
\left[ \bar u(p_2') (-ie) \Gamma^\nu (\ell, p_2, p_2' \vert k) u(p_2) \right]
\delta^{(4)} (p_2 + \ell k - p_2' - Q) ,
\end{align} 
where ${\cal D}_{\mu \nu}$ is the free photon propagator.
Executing the $Q$ integration with one of the $\delta^{(4)}$ distributions  yields
\begin{align}
S_{\mathrm{nlM}} &= \frac{- e^2}{(2 \pi)^6} \int \frac{d \ell}{2 \pi} \frac{d r}{2 \pi}
\delta^{(4)} (p_1 + p_2 + [r + \ell] k - p_1' - p_2') \nonumber\\
&\times
\left[ \bar u(p_1') \Gamma^\mu (r, p_1, p_1' \vert k) u(p_1) \right]
{\cal D}_{\mu \nu} (Q)
\left[ \bar u(p_2') \Gamma^\nu (\ell, p_2, p_2' \vert k) u(p_2) \right] , \label{eq_nlM1}
\end{align} 
where $Q = p_1 + rk - p_1' = p_2' - \ell k - p_2$.
The intermediate photon's virtuality is defined by
$\nu := Q^2$, which is related to the photon number parameters 
\begin{align}
r(\nu) = \frac{\nu - \delta p_1^2}{2 \delta p_1 \cdot k}, 
\quad
\ell (\nu) = \frac{\nu - \delta p_2^2}{2 \delta p_2 \cdot k}, 
\end{align}
with $\delta p_n := p_n - p_n'$, $n = 1, 2$. Thus, the photon number parameters $\ell$ and $r$ 
are intervened. With aid of light cone variables, the four-momentum balance can be rewritten as
\begin{align}
\delta^{(4)} (p_1 + p_2 + [r + \ell] k - p_1' - p_2') &= \delta^{lf} (p_1 + p_2 - p_1' - p_2' ) \nonumber \\ 
 &\times \delta(p_1^+ + p_2^+ + [r + \ell] k^+ - {p_1'}^+ - {p_2'}^+)
\end{align} 
to execute the $\ell$ integral in Eq.~(\ref{eq_nlM1}) with the result
\begin{align}
S_{\mathrm{nlM}} &= \frac{- e^2}{(2 \pi)^6 k^+} 
\delta^{lf} (p_1 + p_2 - p_1' - p_2') \nonumber\\
& \times \int dr
\left[ \bar u(p_1') \Gamma^\mu (r, p_1, p_1' \vert k) u(p_1) \right]
{\cal D}_{\mu \nu} (Q)
\left[ \bar u(p_2') \Gamma^\nu (\ell_r, p_2, p_2' \vert k) u(p_2) \right] , \label{eq_nlM2}
\end{align}
where $\ell_r := r_0 - r$ with $r_0 = (p_1^+ + p_2^+ - {p_1'}^+ - {p_2'}^+) / k^+$ or
\begin{align}
\ell_r = \frac{(p_1' + p_2' - p_2)^2 - m^2 }{2 k \cdot p_1 } - r .
\end{align}

The decomposition (\ref{eq:Gamma_decomp}) facilitates four contributions to the direct term:
\begin{align}
\underbrace{\includegraphics[width=0.22\textwidth,valign=c]{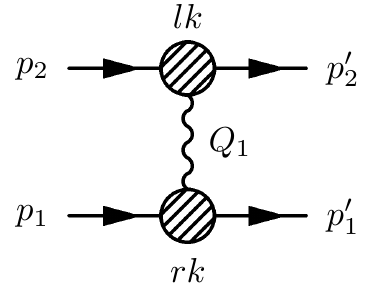}}_{S_{\mathrm{fi}}^{\mathrm{d}}}
&=
\underbrace{\includegraphics[width=0.22\textwidth,valign=c]{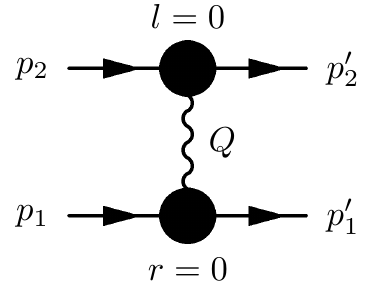}}_{=:S_0}+
\underbrace{\includegraphics[width=0.22\textwidth,valign=c]{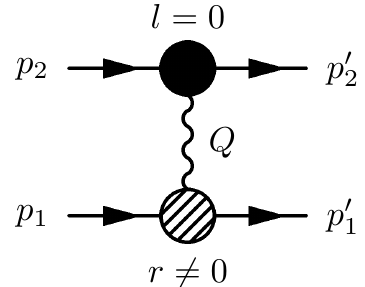}}_{=:S_{11}}+
\underbrace{\includegraphics[width=0.22\textwidth,valign=c]{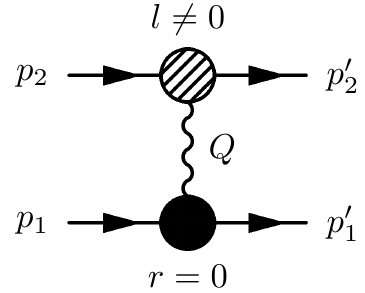}}_{=:S_{12}} \nonumber \\
&+
\underbrace{\includegraphics[width=0.22\textwidth,valign=c]{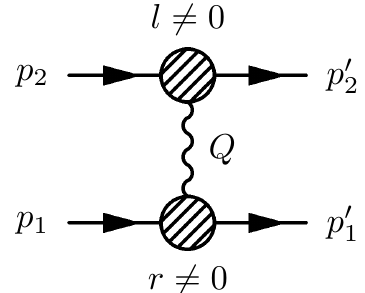}}_{=:S_{2}} .
\label{eq:sf_moller_decomp_diag}
\end{align}
Clearly, w.r.t.\ Eq.~(\ref{eq:Gamma_div}),
the black-bullet vertices do dependent on the background field
since \includegraphics[width=0.12\textwidth,valign=c]{sf_mom_vertex_inf.pdf} 
$\propto \mathcal{G} = e^{iG_+} + e^{i G_-}$ and
$G_\pm = \alpha_1^\mu \int_{\phi_0}^{\pm \infty} d \phi A_\mu(\phi)
+ \alpha_2  \int_{\phi_0}^{\pm \infty} d \phi A(\phi)^2$,
meaning that the first diagram, $S_0$, links to the M{\o}ller scattering in vacuum with momentum balance
$p_1 + p_2 = p_1' + p_2'$, thus reproducing standard perturbative QED.
The background field, in particular, its temporal shape encoded in $g(\phi)$, enters the other three diagrams.
A suggestive interpretation is proposed in \cite{Ilderton:2020rgk}-Fig.~2 by attributing a temporal ordering
to the diagrams. For instance,
the second diagram, $S_{11}$, would refer to virtual Compton under the influence of the background field
(corresponding to the hatched vertex $\Gamma_{\mathrm{reg}}^\mu$),
while the subsequent virtual-photon absorption in the black vertex $\Gamma_{\mathrm{div}}^\mu$
would proceed after the impact of the external field. Such an interpretation would ascribe the first diagram
to proceeding before or after the action of the external field, while the last diagram would refer to both sub-processes
within the action of the field; the third diagram, $S_{12}$, would be accordingly interpreted as virtual Compton prior to
the external impact. Independent of such an interpretation, the fourth diagram, $S_2$,
facilitates on- and off-shell contributions. 

Analog to the sequence of steps in elaborating the matrix elements in Section \ref{sect:nlC}     
one can easily explicate the above diagrams to obtain the decomposition of the four-point amplitude
corresponding to Eq.~(2.23) in \cite{Ilderton:2020rgk}. 
As pointed out in Section \ref{sec:soft_background},
in the case of soft interactions with the background field, 
the finite ($\Gamma^\mu_{\mathrm{reg}}$)
and gauge restoration ($\Gamma^\mu_{\mathrm{div}}$) parts of the dressed vertices factorize 
into a hard scattering part and a generalized soft factor. Consequently, 
this soft/hard factorization also sets in for each diagram in the decomposition \eqref{eq:sf_moller_decomp_diag}. 
Therefore, in the simultaneous limit \( a_0\to 0 \) and \( \Delta \phi \to \infty \), 
the corresponding soft versions of the five-point functions in perturbative monochromatic QED appear, 
where soft photons couple to each Fermion line.

Considering again monochromatic plane-wave background fields and inserting (\ref{eq:Gamma_IPW})
in the M{\o}ller matrix element (\ref{eq_nlM2}), we arrive at 
\begin{align}
\mathcal{S}_{\mathrm{nlM}}^{\textrm{IPW}} &= \frac{- e^2}{(2 \pi)^6 k^+} \sum_{n, n'} \int dr \,
\delta (r - \beta(p_1, p_1', k )- n) \, \delta (\ell_r - \beta(p_2,p_2', k) - n') \\
\times & [ \bar u(p_1') \Gamma_{\textrm{IPW, n}}^\mu (r,p_1, p_1') u(p_1) ]
\mathcal{D}_{\mu \nu} (Q) [\bar u(p_2')  \Gamma_{\textrm{IPW, n'}}^\mu (\ell_r,p_2, p_2') u(p_2) ] .
\end{align}
Solving the $r$-integral, we find $r = \beta(p_1, p_1',k) + n$ and, therefore, the virtuality reads
$\nu (r = \beta + n) = (\beta(p_1, p_1', k) + n) 2 (p_1 - p_1') \cdot k +(p_1 -p_1')^2$
meeting the resonance condition $\nu = 0$ reveals similar Oleinik resonances as shown in sub-section
\ref{sec:Oleinik}.

Via crossing symmetry, the trident amplitude 
\includegraphics[width=0.12\textwidth,valign=c]{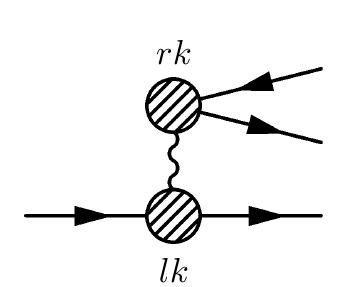}
has a similar decomposition as given above by the four direct-term
diagrams, to be supplemented by the exchange terms. The changed $in$- and $out$-phase space, however,
modifies the treatment/interpretation of individual contributions,
to be dealt with in a follow-up paper. 
The nonlinear M{\o}ller scattering is laser-assisted, while the nonlinear trident is laser-enabled.
Oleinik resonances show up in both channels \cite{Sizykh:2021ywt}.

\section{Summary}\label{sect:Summary}

Following \cite{HernandezAcosta:2019vok,Acosta:2020dud} we present the comprehensive
momentum-space Furry-picture Feynman rules for QED in an external classical
background field. Our emphasis is on formal aspects of 
gauge invariance and Ward identity,
thus providing a general framework well-suited for $n$-point amplitudes.
The special case of four-point amplitudes, dealt with in \cite{Ilderton:2020rgk}
within a somewhat different formulation, emerges naturally.
Three-point amplitudes are considered exhaustively in the past and are
uncovered as well. The benefit of our formalism is a systematic approach to
the weak-field approximation, i.e.\ the series expansion of the $S$ matrix element in powers of the
laser intensity parameter $a_0$. The leading-order term represents 
``pulsed perturbative QED" which accounts, in contrast to standard 
perturbative QED, for the temporal structure of the external classical field,
where the Fourier transform of that field enters decisively.
The limiting case of a monochromatic external classical field recovers
the standard perturbative QED in terms of Feynman graphs and their rules
of translation into amplitudes for processes with one incoming photon
impinging on a target, e.g.\ an electron or a scattering 
electron-electron/positron system.
The next-to-leading order terms in $a_0$
are determined by Fourier transforms of the external classical field
in various (nonlinear) combinations. The monochromatic limiting case
describes processes with two incoming photons impinging on the target, 
thus referring to the two-photon channel, 
e.g.\ $k_1 + k_2 + e^- \to X$, with $X = {e^-}' e^+ e^-$
for trident. When considering one monochromatic laser field,
$k_1 = k_2$. One may also deal with $k_1 \ne k_2$, e.g.\ for the
superposition of optical laser and XFEL beams.
We leave the explication of such processes w.r.t.\ trident for separate work.     

Elements of our QED momentum-space Furry-picture Feynman rules are 
free Dirac spinors,  free Fermion propagator, and free photon propagator,
and the external field impact is solely encoded in the Fermion-Fermion-photon vertex function.    
By providing a suitable framework for the evaluation of the latter vertex function,
standard platforms for the calculation of Feynman diagrams can be used for strong-field QED processes.

\begin{appendix}

\section{Ward identity and regularization of $\mathbf{B_0}$ \label{sect:Ward_ID}}

In the case of an external photon $K$ attached to the vertex $\Gamma$, the condition
$K \cdot \Gamma = 0$ ensures the independence on the arbitrary gauge function $q(K)$,
when the polarization four vector is gauged according to $\epsilon \to \epsilon' + q K \cdot \Gamma$.
Analogously, the photon propagator 
${\cal D}_{\mu \nu} = 
\frac{i}{K^2 + i \epsilon} \left( - \eta_{\mu \nu} + (1 - \xi) \frac{K\mu K_\nu}{K^2 - i \epsilon} \right)$,
connecting two adjacent vertices does not leave any dependence of the gauge parameter $\xi$ on
the structure $\Gamma \cdot {\cal D} \cdot \Gamma$ if the same Ward identity $K \cdot \Gamma = 0$ 
is fulfilled. Sandwiching the Ward identity by Dirac spinors yields, for $K = k'$,\footnote{
Strictly speaking, this treatment refers essentially to on-shell Fermions. However, in gauge theories more general identities appear, known as Ward-Takahashi identities,
where some of the external Fermions are off-shell. Such relations require further considerations
w.r.t.\ dressed vertex.}  
\begin{equation} \label{eq:Ward}
0 = \overline u_{p'}k' _\mu\Gamma ^\mu u_p
=  \left( \overline u_{p'}k' _\mu \gamma ^\mu u_p\right)B_0 +  \left( \overline u_{p'}k' _\mu\Gamma ^{\mu \nu}_1 u_p\right) B_{1 \nu}+\left( \overline u_{p'}k' _\mu\Gamma ^{\mu}_2 u_p\right) B_2,
\end{equation}
where $\{B_0, B_{1 \nu}, B_2\}$ denote the phase integrals (\ref{eq:B0}, \ref{eq:B1}, \ref{eq:B2}) 
and $\{ \gamma ^\mu _0, \Gamma ^{\mu \nu} _1, \Gamma_2 ^\mu \}$ are the elementary vertices 
(\ref{eq:vertex1}, \ref{eq:vertex2}) (for both we suppress the momentum dependence for now). 
The free Dirac spinors in the side-condition \eqref{eq:Ward}
appear either if the vertex is attached by $in/out$ Fermion lines or by propagators, 
where the spin sum decomposition \( \slashed p + m = \sum_ \sigma u_{ \sigma p} \overline u_{ \sigma p} \) (analog for the other Dirac spinors) can be used. The energy-momentum balance $p + \ell k = p' + k'$ holds at each vertex, 
which is implied by the delta distribution in the fully dressed vertex $\Gamma$, 
the Ward identity \eqref{eq:Ward} of the dressed vertex function reads 
$0 = \left( \ell B_0(\ell) + \alpha_1 ^\mu B_{1 \mu}(\ell) + \alpha_2 B_2(\ell)\right) (\overline u_{p'} \slashed k u_p)$
upon using  Dirac equations in momentum space,  $(\slashed p - m) u_p = 0$ and $\overline u_{p'}(\slashed p' - m) = 0$, respectively. This implies a severe constraint for the phase integrals:
\begin{equation}\label{eq:Ward2}
0 = \ell B_0(\ell) + \alpha_1 ^\mu B_{1 \mu}(\ell) + \alpha_2 B_2(\ell),
\end{equation}
being equivalent to the Ward identity \eqref{eq:Ward} and shows that the phase integrals 
$\{B_0, B_{1 \nu}, B_2\}$ are not independent.
The relation \eqref{eq:Ward2} is a well-known formula, which appears in several investigations of specific processes 
in strong-field QED, e.g.\ \cite{Seipt:2012nad} in the case of Compton scattering 
or for the trident process \cite{Ilderton:2010wr}. However, the connection to the Ward identity and, therefore, 
to gauge invariance was not stressed there. 

The Ward identity \eqref{eq:Ward} must be solved for one of the phase integrals, e.g.\ $ B_0$, 
in a distributional manner. This can be formulated as follows. Let be $ b_0(\ell)$ a solution of Eq.~(\ref{eq:Ward2}), 
i.e.\ $0 = \ell b_0(\ell) + \alpha_1 ^\mu B_{1 \mu}(\ell) + \alpha_2 B_2(\ell)$, 
then $\hat b_0(\ell) := b_0(\ell) + \mathcal{G} \delta(\ell) \) is a solution as well, 
where $ \mathcal{G}$ is an arbitrary, but finite, function of the momenta. This seems trivial, 
because $\ell \delta(\ell) = 0$, but it turns out that this term leads to non-negligible contributions. 
However, the Ward identity \eqref{eq:Ward2} does not determine the delta distribution's prefactor $\mathcal{G}$.
Instead, it can be derived by regulating the integral
in the definition of $B_0$, Eq.~(\ref{eq:B0}).
Adapting the procedure in
\cite{Boca:2012pz} for Compton scattering in a generic case by inserting
$e^{- \epsilon \vert \phi \vert}$ with $\epsilon >0$ in the integral in Eq.~(\ref{eq:B0}) we get
\begin{align}
B_0(\ell)&=\lim_{ \epsilon\to 0^+} \int_{-\infty}^{\infty}\mathrm{d} \phi\, e^{- \epsilon 
\vert \phi \vert}e^{i \ell \phi} e^{iG( \phi)}\\
&=\lim_{ \epsilon\to 0^+} \bigg[\int_{-\infty}^{0}\mathrm{d} \phi\,e^{(i \ell + \epsilon) \phi} e^{iG( \phi)}
+ \int_{0}^{\infty}\mathrm{d} \phi\, e^{(i \ell- \epsilon) \phi} e^{iG( \phi)}\bigg] \nonumber\\
&\phantom{=\lim_{ \epsilon\to 0^+}\bigg[}+ \frac{e^{(i \ell + \epsilon) \phi} e^{iG( \phi)}}{i \ell + \epsilon}
\bigg\vert_{- \infty}^{0} - \frac{i}{i \ell - \epsilon}
\int_{- \infty}^{0}\mathrm{d} \phi\,e^{(i \ell - \epsilon) \phi}G'( \phi) e^{iG( \phi)}\bigg] 
\nonumber \\
&\phantom{=\lim_{ \epsilon\to 0^+}\bigg[}+ \frac{e^{(i \ell - \epsilon) \phi} e^{iG( \phi)}}{i \ell - \epsilon}\bigg\vert_{0}^{\infty} - \frac{i}{i \ell - \epsilon}\int_{0}^{\infty}\mathrm{d} \phi\,e^{(i \ell - \epsilon) \phi}G'( \phi) e^{iG( \phi)}\bigg]\label{gl:B0partialInt},
\end{align}
where we use partial integration and the shortcut \( G':=\frac{\mathrm{d}}{\mathrm{d} \phi}G \). 
Considering the non-integral terms of $B_0$, one gets
\begin{align}
\lim_{ \epsilon\to 0^+} \left[ \frac{e^{(il \ell+ \epsilon) \phi} 
e^{iG( \phi)}}{i \ell + \epsilon}\bigg\vert_{- \infty}^{0} +  \frac{e^{(i \ell - \epsilon) \phi} e^{iG( \phi)}}{i \ell - \epsilon}\bigg\vert_{0}^{\infty} \right] = 2\lim_{ \epsilon\to 0^+}  \left[ \frac{ \epsilon}{ \epsilon^2 + \ell^2}\right] e^{i G(0)}
= 2 \pi \delta(\ell) e^{i G(0)} .
\label{gl:B0deltaPart1}
\end{align}
In the very last step, we perform the limit in a distributional manner.
To evaluate the other terms in Eq.~(\ref{gl:B0partialInt}), we consider the integral 
$\int_{-\infty}^{0}\mathrm{d} \phi\,e^{(i \ell + \epsilon) \phi}G'( \phi) e^{iG( \phi)}$ 
which is finite for every $\epsilon \geq 0$ due to the proportionality $G'(\phi) \sim A ^\mu (\phi)$
for all $\phi_0$,
where $A ^\mu$ is assumed to vanish at the lower limit of the integral,
i.e.\ $\lim_{\phi \to - \infty} = 0$.
The same holds for the other integral, so the limit in these integrals can be executed to get
\begin{align}
&\lim_{ \epsilon\to 0^+} \bigg[\frac{i}{i \ell + \epsilon}\int_{-\infty}^{0}\mathrm{d} \phi\,e^{i \ell \phi}G' e^{iG} 
+ \frac{i}{i \ell - \epsilon}\int_{0}^{\infty}\mathrm{d} \phi\,e^{i \ell \phi}G'e^{iG}\bigg]\\
&= \lim_{ \epsilon\to 0^+} \frac{1}{2}\bigg[ \left(\frac{i}{i \ell + \epsilon} +  
\frac{i}{i \ell - \epsilon} \right)\int_{-\infty}^{\infty}\mathrm{d} \phi\,
e^{i \ell \phi}G' e^{iG}\nonumber\\
&\phantom{= \lim_{ \epsilon\to 0^+}\frac{1}{2}}+ \left(\frac{i}{i \ell + \epsilon} -  
\frac{i}{i \ell - \epsilon} \right)\left(\int_{-\infty}^{0}\mathrm{d} \phi\,
e^{i \ell \phi}G' e^{iG} -\int_{0}^{\infty}\mathrm{d} \phi\,
e^{i \ell \phi}G'e^{iG} \right)\bigg],\label{gl:secondLimitB0}
\end{align}
where we apply the identity $uw + vz = \frac{ u+v}{2}(w+z) + \frac{ u-v}{2} \left(w-z\right)$ 
with $\{u,v,w,z\} \in \mathbb C$. Starting with the first term in Eq.~(\ref{gl:secondLimitB0}), we get 
\begin{align}
 \lim_{ \epsilon\to 0^+} \frac{1}{2} \left(\frac{i}{i \ell + \epsilon} +  \frac{i}{i \ell - \epsilon} \right)\int_{-\infty}^{\infty}\mathrm{d} \phi\,e^{i \ell \phi}G' e^{iG}= \mathcal P \left[ \frac{1}{\ell}\int_{-\infty}^{\infty}\mathrm{d} \phi\,
e^{i \ell \phi}G' e^{iG} \right]
\end{align}
by using
\begin{align}
\lim_{ \epsilon\to 0^+}\int_{a}^b \mathrm{d}x \frac{x^2}{x^2+ \epsilon^2} H(x)
 = \mathcal P\int_{a}^b \mathrm{d}x \, H(x),
\end{align}
with an arbitrary function \( H:(a,b)\to \mathbb C\). 
The second term in Eq.~(\ref{gl:secondLimitB0}) contains again a delta-distribution:
\begin{align}
\lim_{ \epsilon\to 0^+} &\frac{1}{2}\left(\frac{i}{i \ell + \epsilon} -  \frac{i}{i \ell - \epsilon} \right)\left(\int_{-\infty}^{0}\mathrm{d} \phi\,e^{i \ell \phi}G' e^{iG} -\int_{0}^{\infty}\mathrm{d} \phi\,
e^{i \ell \phi}G'e^{iG} \right)\\
&=i \pi \delta(\ell)\left(\int_{-\infty}^{0}\mathrm{d} \phi\,
e^{i \ell \phi}G' e^{iG} -\int_{0}^{\infty}\mathrm{d} \phi\,e^{i \ell \phi}G'e^{iG} \right)\\\, 
&=i \pi \delta(\ell)\left(\int_{-\infty}^{0}\mathrm{d} \phi\,G' e^{iG} -\int_{0}^{\infty}\mathrm{d} \phi\,G'e^{iG} \right),
\end{align}
where we used \( \delta(x)H(x)= \delta(x)H(0)\) in the last step. 
To evaluate these integrals, we use $G'e^{iG}=-i \left( e^{iG}\right)'$:
\begin{align} \label{eq:(13)}
\int_{-\infty}^{0}\mathrm{d} \phi\,G' e^{iG} -\int_{0}^{\infty}\mathrm{d} \phi\,G'e^{iG}&= \frac{1}{i}\left( e^{iG}\bigg\vert_{-\infty}^0 -e^{iG} \bigg\vert_{0}^{\infty}\right)= \frac{1}{i}\left( 2 e^{i G(0)} - e^{iG_{+}} - e^{iG_{-}}\right)
\end{align}
with the abbreviation $G_{\pm} := \lim_{ \phi\to \infty} G(\pm \phi)$. Finally, we insert Eqs.~(\ref{gl:B0deltaPart1}) and (\ref{gl:secondLimitB0}) in Eq.~(\ref{gl:B0partialInt}) and repeat the evaluation as above, to obtain
\begin{align}
B_0(\ell) =  \left( e^{iG_{+}} + e^{iG_{-}} \right) \pi \delta (\ell) - 
\mathcal P\left[ \frac{1}{\ell}\int_{-\infty}^{\infty}\mathrm{d} \phi\,e^{i \ell \phi}G' e^{iG} \right].
\end{align}
Note the independence on both, $G(0)$ and $\phi_0$
as a consequence of Eqs.~(\ref{gl:B0partialInt}, \ref{gl:B0deltaPart1}, \ref{eq:(13)}).
Considering $G'( \phi) = \alpha _1 ^\mu A _\mu(\phi) + \alpha_2 A^2(\phi)$ (cf.\ Eq.~(\ref{eq:G}))
and the definitions of $B_1 ^\mu$ as well as $B_2$ in Eq.~(\ref{eq:B1}, \ref{eq:B2}), respectively, 
we can write $B_0$ in terms of the other phase integrals as 
\begin{align}\label{eq:phaseintegral0_reg}
B_0 (\ell)&= \pi \delta(\ell) \left( e^{iG_+} + e^{iG_-}\right) - 
\mathcal P \left[ \frac{ \alpha_1 ^\mu }{\ell} B_{1 \mu}(\ell) + \frac{ \alpha_2}{\ell}B_2(\ell)\right]\\
&=\pi \delta(\ell) \mathcal G + \hat B_0(\ell),
\end{align}
where we introduce the abbreviation 
\begin{align}
\mathcal G(p,p',k) := \exp \left( i G^+(p,p',k)\right) + \exp \left( i G^-(p,p',k)\right),\label{eq:div_prefactor}
\end{align}
with the asymptotic values $G^\pm$ (\ref{eq:Gpm})
of the non-linear phase \eqref{eq:G}
as well as the finite phase integral 
$\hat B_0(l) := - \mathcal P \left[  \frac{\alpha_1^\mu}{\ell}  B_{1 \mu}(\ell) + \frac{\alpha_2}{\ell} B_2(\ell)\right]$. 
Finally, it is easy to see that the regularized version \eqref{eq:phaseintegral0_reg} of the phase integral solves 
Eq.~(\ref{eq:Ward}), which implies 
the solution of the Ward identity Eq.~(\ref{eq:Ward2}). 
Inserting the regularized version of the phase integral \eqref{eq:phaseintegral0_reg}, 
the dressed vertex function (\ref{eq:vertex}) decomposes as
$\Gamma^\mu(\ell, p,p' \vert k) = \Gamma^\mu_{\mathrm{div}}(\ell, p,p'\vert k) 
+ \Gamma^\mu_{\mathrm{reg}}(\ell, p,p'\vert k)$, 
where the divergent part is given by Eq.~(\ref{eq:Gamma_div})
and the finite part is given by  Eq.~(\ref{eq:Gamma_reg}).

One may interpret these parts of the dressed vertex as follows. The divergent part $\Gamma^\mu_{\mathrm{div}}$, 
Eq.~(\ref{eq:Gamma_div}), enforces $\ell=0$. 
Therefore, it can be considered as part of the dressed vertex function with no momentum exchange with the background field. 
This has no contribution to one-vertex processes like nonlinear Compton scattering or nonlinear Breit-Wheeler pair production due to the vanishing physical phase space, 
i.e.\ there is neither single-photon absorption nor single-photon emission in perturbative QED. 
However, for processes with more than one vertex, e.g.\ the trident process, 
the vanishing momentum exchange from the background field to one vertex 
may eventually be compensated due to the momentum transfer at another vertex. 
Since the only dependence of this non-transfer term on the background field is condensed in the factor $\mathcal G$, 
the leading order in $A ^\mu$ of the whole non-vanishing term is constant through 
$\mathcal G = 2 + \mathcal O(A ^\mu )$. 

The finite part $\Gamma_{reg}^\mu$,
Eq.~(\ref{eq:Gamma_reg}), may be interpreted as a part of the dressed vertex function 
with a genuine momentum transfer from the background field to the vertex, 
which is indicated by the occurring principal value in the finite part $\hat B_0(\ell)$ 
of the regularized phase integral \eqref{eq:phaseintegral0_reg}, considering the other phase integrals are regular for $\ell \to 0$. 
Moreover, since the elementary vertices (\ref{eq:vertex1}, \ref{eq:vertex2}) 
as well as the kinematic factors $\alpha_i$ (\ref{eq:alpha12}) 
are independent of the background field, the leading order of the finite part $\Gamma ^\mu _{\mathrm{reg}}$ 
of the vertex function $\Gamma ^\mu$ is linear in $A ^\mu$, i.e.\ there is no $A ^\mu$-independent term 
in an expansion of $\Gamma ^\mu$ w.r.t.\ the background field. 

\end{appendix} 

\begin{acknowledgements}

We gratefully acknowledge our former collaboration with
D.~Seipt, T.~Nousch, A.~Otto, A.I.~Titov, and T.~Heinzl
on various topics of strong-field QED
as well as 
R.~Sauerbrey, T.E.~Cowan, U.~Schramm, and H.P.~Schlenvoigt
w.r.t.\ HIBEF.
Useful discussions with R.~Sch\"utzhold, G.~Torgrimsson, C.~Kohlf\"urst, and N.~Ahmadiniaz
are thanked for. 

The work of UHA was partly funded by the Center for Advanced Systems Understanding (CASUS) that is financed by Germany’s Federal Ministry of Education and Research (BMBF) and by the Saxon Ministry for Science, Culture and Tourism (SMWK) with tax funds on the basis of the budget approved by the Saxon State Parliament.

\end{acknowledgements}

{}


\begin{thebibliography}{99}

\bibitem{world_record}
J. Yoon, Y. Kim, I. Choi, J. Sung, H. Lee, S. Lee, and C. Nam, 
``Realization of laser intensity over $10^{23}$ W/cm${}^2$," 
Optica \textbf{8}, 630-635 (2021).

\bibitem{Marklund:2022gki}
M.~Marklund, T.~G.~Blackburn, A.~Gonoskov, J.~Magnusson, S.~S.~Bulanov and A.~Ilderton,
``Towards critical and supercritical electromagnetic fields,''
[arXiv:2209.11720 [physics.plasm-ph]].

\bibitem{Ilderton:2021zej}
A.~Ilderton,
``Physics of adiabatic particle number in the Schwinger effect,''
Phys. Rev. D \textbf{105}, no.1, 016021 (2022)
[arXiv:2108.13885 [hep-ph]].

\bibitem{Kohlfurst:2021skr}
C.~Kohlf\"urst, N.~Ahmadiniaz, J.~Oertel and R.~Sch\"utzhold,
``Sauter-Schwinger Effect for Colliding Laser Pulses,''
Phys. Rev. Lett. \textbf{129}, no.24, 241801 (2022)
[arXiv:2107.08741 [hep-ph]].

\bibitem{Sevostyanov:2020dhs}
D.~G.~Sevostyanov, I.~A.~Aleksandrov, G.~Plunien and V.~M.~Shabaev,
``Total yield of electron-positron pairs produced from vacuum in strong electromagnetic fields: Validity of the locally constant field approximation,''
Phys. Rev. D \textbf{104}, no.7, 076014 (2021)
[arXiv:2012.10751 [hep-ph]].

\bibitem{Heinzl:2021mji}
T.~Heinzl, A.~Ilderton and B.~King,
``Classical Resummation and Breakdown of Strong-Field QED,''
Phys. Rev. Lett. \textbf{127}, no.6, 061601 (2021)
[arXiv:2101.12111 [hep-ph]].

\bibitem{Taya:2020dco}
H.~Taya, T.~Fujimori, T.~Misumi, M.~Nitta and N.~Sakai,
``Exact WKB analysis of the vacuum pair production by time-dependent electric fields,''
JHEP \textbf{03}, 082 (2021)
[arXiv:2010.16080 [hep-th]].

\bibitem{Edwards:2020npu}
J.~P.~Edwards and A.~Ilderton,
``Resummation of background-collinear corrections in strong field QED,''
Phys. Rev. D \textbf{103}, no.1, 016004 (2021)
[arXiv:2010.02085 [hep-ph]].

\bibitem{Ilderton:2019kqp}
A.~Ilderton,
``Note on the conjectured breakdown of QED perturbation theory in strong fields,''
Phys. Rev. D \textbf{99}, no.8, 085002 (2019)
[arXiv:1901.00317 [hep-ph]].

\bibitem{Fedotov:2016afw}
A.~M.~Fedotov,
J. Phys. Conf. Ser. \textbf{826}, no.1, 012027 (2017)
[arXiv:1608.02261 [hep-ph]].

\bibitem{Adamova:2019vkf}
D.~Adamov\'a, G.~Aglieri Rinella, M.~Agnello, Z.~Ahammed, D.~Aleksandrov, A.~Alici, A.~Alkin, T.~Alt, I.~Altsybeev and D.~Andreou, \textit{et al.}
``A next-generation LHC heavy-ion experiment,''
[arXiv:1902.01211 [physics.ins-det]].

\bibitem{Strominger:2017zoo}
A.~Strominger,
``Lectures on the Infrared Structure of Gravity and Gauge Theory,''
[arXiv:1703.05448 [hep-th]].

\bibitem{Kapec:2015ena}
D.~Kapec, M.~Pate and A.~Strominger,
``New Symmetries of QED,''
Adv. Theor. Math. Phys. \textbf{21}, 1769-1785 (2017)
[arXiv:1506.02906 [hep-th]].

\bibitem{Feal:2022iyn}
X.~Feal, A.~Tarasov and R.~Venugopalan,
``QED as a many-body theory of worldlines: General formalism and infrared structure,''
Phys. Rev. D \textbf{106}, no.5, 056009 (2022)
[arXiv:2206.04188 [hep-th]].

\bibitem{Fedotov:2022ely}
A.~Fedotov, A.~Ilderton, F.~Karbstein, B.~King, D.~Seipt, H.~Taya and G.~Torgrimsson,
``Advances in QED with intense background fields,''
Phys. Rept. \textbf{1010}, 1-138 (2023)
[arXiv:2203.00019 [hep-ph]].

\bibitem{Gonoskov:2021hwf}
A.~Gonoskov, T.~G.~Blackburn, M.~Marklund and S.~S.~Bulanov,
``Charged particle motion and radiation in strong electromagnetic fields,''
Rev. Mod. Phys. \textbf{94}, no.4, 045001 (2022)
[arXiv:2107.02161 [physics.plasm-ph]].

\bibitem{DiPiazza:2011tq}
A.~Di Piazza, C.~M\"uller, K.~Z.~Hatsagortsyan and C.~H.~Keitel,
``Extremely high-intensity laser interactions with fundamental quantum systems,''
Rev. Mod. Phys. \textbf{84}, 1177 (2012)
[arXiv:1111.3886 [hep-ph]].

\bibitem{Ilderton:2017xbj}
A.~Ilderton and D.~Seipt,
``Backreaction on background fields: A coherent state approach,''
Phys. Rev. D \textbf{97}, no.1, 016007 (2018)
[arXiv:1709.10085 [hep-th]].

\bibitem{Seipt:2016fyu}
D.~Seipt, T.~Heinzl, M.~Marklund and S.~S.~Bulanov,
``Depletion of Intense Fields,''
Phys. Rev. Lett. \textbf{118}, no.15, 154803 (2017)
[arXiv:1605.00633 [hep-ph]].

\bibitem{HernandezAcosta:2019vok}
U.~Hernandez Acosta and B.~K\"ampfer,
``Laser pulse-length effects in trident pair production,''
Plasma Phys. Control. Fusion \textbf{61}, no.8, 084011 (2019)
[arXiv:1901.08860 [hep-ph]].

\bibitem{Ilderton:2020rgk}
A.~Ilderton and A.~J.~MacLeod,
``The analytic structure of amplitudes on backgrounds from gauge invariance and the infra-red,''
JHEP \textbf{04}, 078 (2020)
[arXiv:2001.10553 [hep-th]].


\bibitem{Agarwal:2021ais}
N.~Agarwal, L.~Magnea, C.~Signorile-Signorile and A.~Tripathi,
``The infrared structure of perturbative gauge theories,''
Phys. Rept. \textbf{994}, 1-120 (2023)
[arXiv:2112.07099 [hep-ph]].

\bibitem{Acosta:2020dud}
U.~H.~Acosta,
``Pulsed-perturbative QED: A study of trident pair production in pulsed laser fields,''
PhD thesis, TU Dresden 2020.

\bibitem{Ritus85}
  V. I. Ritus,
 ``Quantum effects of the interaction of elementary particles
with an intense electromagnetic field”.
 J. Sov. Laser Res. (United States), \textbf{6:5}, 497 (1985).

\bibitem{Meuren:2013oya}
S.~Meuren, C.~H.~Keitel and A.~Di Piazza,
``Polarization operator for plane-wave background fields,''
Phys. Rev. D \textbf{88}, no.1, 013007 (2013)
[arXiv:1304.7672 [hep-ph]].

\bibitem{Seipt:2016rtk}
D.~Seipt, V.~Kharin, S.~Rykovanov, A.~Surzhykov and S.~Fritzsche,
``Analytical results for nonlinear Compton scattering in short intense laser pulses,''
J. Plasma Phys. \textbf{82}, no.2, 655820203 (2016)
[arXiv:1601.00442 [hep-ph]].

\bibitem{Peskin:1995ev}
M.~E.~Peskin and D.~V.~Schroeder,
``An Introduction to quantum field theory,''
Addison-Wesley, 1995,
ISBN 978-0-201-50397-5

\bibitem{Weinberg:1995mt}
S.~Weinberg,
``The Quantum theory of fields. Vol. 1: Foundations,''
Cambridge University Press, 2005,
ISBN 978-0-521-67053-1, 978-0-511-25204-4

\bibitem{Seipt:2012nad}
D.~Seipt,
``Strong-Field QED Processes in Short Laser Pulses,''
PhD thesis TU Dresden (2012).

\bibitem{Seipt:2013hda}
D.~Seipt and B.~K\"ampfer,
``Laser assisted Compton scattering of X-ray photons,''
Phys. Rev. A \textbf{89}, no.2, 023433 (2014)
[arXiv:1309.2092 [physics.atom-ph]].

\bibitem{Seipt:2015rda}
D.~Seipt, A.~Surzhykov, S.~Fritzsche and B.~K\"ampfer,
``Caustic structures in x-ray Compton scattering off electrons driven by a short intense laser pulse,''
New J. Phys. \textbf{18}, no.2, 023044 (2016)
[arXiv:1507.08868 [hep-ph]].

\bibitem{Seipt:2019yds}
D.~Seipt, V.~Y.~Kharin and S.~G.~Rykovanov,
``Optimizing Laser Pulses for Narrow-Band Inverse Compton Sources in the High-Intensity Regime,''
Phys. Rev. Lett. \textbf{122}, no.20, 204802 (2019)
[arXiv:1902.10777 [physics.plasm-ph]].

\bibitem{Valialshchikov:2022ndd}
M.~A.~Valialshchikov, D.~Seipt, V.~Y.~Kharin and S.~G.~Rykovanov,
``Towards high photon density for Compton Scattering by spectral chirp,''
Phys. Rev. A \textbf{106}, L031501 (2022)
[arXiv:2204.12245 [physics.acc-ph]].

\bibitem{HernandezAcosta:2020agu}
U.~Hernandez Acosta, A.~Otto, B.~K\"ampfer and A.~I.~Titov,
``Nonperturbative signatures of nonlinear Compton scattering,''
Phys. Rev. D \textbf{102}, no.11, 116016 (2020)
[arXiv:2001.03986 [hep-ph]].

\bibitem{Seipt:2012tn}
D.~Seipt and B.~K\"ampfer,
``Two-photon Compton process in pulsed intense laser fields,''
Phys. Rev. D \textbf{85}, 101701 (2012)
[arXiv:1201.4045 [hep-ph]].

\bibitem{Mackenroth:2012rb}
F.~Mackenroth and A.~Di Piazza,
``Nonlinear Double Compton Scattering in the Ultrarelativistic Quantum Regime,''
Phys. Rev. Lett. \textbf{110}, no.7, 070402 (2013)
[arXiv:1208.3424 [hep-ph]].

\bibitem{Lotstedt:2009zz}
E.~Lotstedt and U.~D.~Jentschura,
``Nonperturbative Treatment of Double Compton Backscattering in Intense Laser Fields,''
Phys. Rev. Lett. \textbf{103}, 110404 (2009)
[arXiv:0909.4984 [quant-ph]].

\bibitem{Loetstedt:2009zz}
E.~Loetstedt and U.~D.~Jentschura,
``Correlated two-photon emission by transitions of Dirac-Volkov states in intense laser fields: QED predictions,''
Phys. Rev. A \textbf{80}, 053419 (2009)
[arXiv:0911.4765 [quant-ph]].

\bibitem{soft_photon_ALICE3}
P.~Braun-Munzinger.
``Soft photons, the Low theorem, and ALICE 3'', EMMI Rapid Reaction Task Force (RRTF), 
\url{https://indico.gsi.de/event/11946/contributions/50402/attachments/34543/45390/2021_RRTF_Low.pdf}

\bibitem{Oleinik:1967}
V.~P.~Oleinik,
``RESONANCE EFFECTS IN THE FIELD OF AN INTENSE LASER RAY",
Sov.\ Phys.\ JETP 
 \textbf{25}, 697 (1967). 

\bibitem{Oleinik:1968}
V.~P.~Oleinik,
``RESONANCE EFFECTS IN THE FIELD OF AN INTENSE LASER RAY. II",
Sov.\ Phys.\ JETP 
\textbf{26}, 1668 (1968). 

\bibitem{Lotstedt:2012zz}
E.~Lotstedt and U.~D.~Jentschura,
``Triple Compton Effect: A Photon Splitting into Three upon Collision with a Free Electron,''
Phys. Rev. Lett. \textbf{108}, 233201 (2012)
[arXiv:1205.0317 [quant-ph]].

\bibitem{Lotstedt:2013uya}
E.~L\"otstedt and U.~D.~Jentschura,
``Theoretical study of the Compton effect with correlated three-photon emission: From the differential cross section to high-energy triple-photon entanglement,''
Phys. Rev. A \textbf{87}, no.3, 033401 (2013)
[arXiv:1405.1669 [quant-ph]].

\bibitem{Dinu:2019pau}
V.~Dinu and G.~Torgrimsson,
``Approximating higher-order nonlinear QED processes with first-order building blocks,''
Phys. Rev. D \textbf{102}, no.1, 016018 (2020)
[arXiv:1912.11015 [hep-ph]].

\bibitem{Dinu:2018efz}
V.~Dinu and G.~Torgrimsson,
``Single and double nonlinear Compton scattering,''
Phys. Rev. D \textbf{99}, no.9, 096018 (2019)
[arXiv:1811.00451 [hep-ph]].

\bibitem{Titov:2020taw}
A.~I.~Titov and B.~Kampfer,
``Non-linear Breit\textendash{}Wheeler process with linearly polarized beams,''
Eur. Phys. J. D \textbf{74}, no.11, 218 (2020)
[arXiv:2006.04496 [hep-ph]].

\bibitem{Titov:2019kdk}
A.~I.~Titov, A.~Otto and B.~Kampfer,
``Multi-photon regime of non-linear Breit-Wheeler and Compton processes in short linearly and circularly polarized laser pulses,''
Eur. Phys. J. D \textbf{74}, no.2, 39 (2020)
[arXiv:1907.00643 [physics.plasm-ph]].

\bibitem{Titov:2013kya}
A.~I.~Titov, B.~K\"ampfer, H.~Takabe and A.~Hosaka,
``Breit-Wheeler process in very short electromagnetic pulses,''
Phys. Rev. A \textbf{87}, no.4, 042106 (2013)
[arXiv:1303.6487 [hep-ph]].

\bibitem{Nousch:2012xe}
T.~Nousch, D.~Seipt, B.~Kampfer and A.~I.~Titov,
``Pair production in short laser pulses near threshold,''
Phys. Lett. B \textbf{715}, 246-250 (2012)

\bibitem{Titov:2012rd}
A.~I.~Titov, H.~Takabe, B.~Kampfer and A.~Hosaka,
``Enhanced subthreshold electron-positron production in short laser pulses,''
Phys. Rev. Lett. \textbf{108}, 240406 (2012)
[arXiv:1205.3880 [hep-ph]].

\bibitem{Titov:2018bgy}
A.~I.~Titov, B.~K\"ampfer and H.~Takabe,
``Nonlinear Breit-Wheeler process in short laser double pulses,''
Phys. Rev. D \textbf{98}, no.3, 036022 (2018)
[arXiv:1807.04547 [physics.plasm-ph]].

\bibitem{Otto:2016fdo}
A.~Otto, T.~Nousch, D.~Seipt, B.~K\"ampfer, D.~Blaschke, A.~D.~Panferov, S.~A.~Smolyansky and A.~I.~Titov,
``Pair production by Schwinger and Breit\textendash{}Wheeler processes in bi-frequent fields,''
J. Plasma Phys. \textbf{82}, no.3, 655820301 (2016)
[arXiv:1604.00196 [hep-ph]].

\bibitem{Nousch:2015pja}
T.~Nousch, D.~Seipt, B.~K\"ampfer and A.~I.~Titov,
``Spectral caustics in laser assisted Breit\textendash{}Wheeler process,''
Phys. Lett. B \textbf{755}, 162-167 (2016)
[arXiv:1509.01983 [physics.plasm-ph]].

\bibitem{Borysov:2022cwc}
O.~Borysov, B.~Heinemann, A.~Ilderton, B.~King and A.~Potylitsyn,
``Using the nonlinear Breit-Wheeler process to test nonlinear vacuum birefringence,''
Phys. Rev. D \textbf{106}, no.11, 116015 (2022)
[arXiv:2209.12908 [hep-ph]].

\bibitem{Ahmadiniaz:2022nrv}
N.~Ahmadiniaz, T.~E.~Cowan, J.~Grenzer, S.~Franchino-Vi\~nas, A.~L.~Garcia, M.~Smid, T.~Toncian, M.~A.~Trejo and R.~Sch\"utzhold,
``Detection schemes for quantum vacuum diffraction and birefringence,''
[arXiv:2208.14215 [physics.optics]].

\bibitem{Ahmadiniaz:2020lbg}
N.~Ahmadiniaz, T.~E.~Cowan, R.~Sauerbrey, U.~Schramm, H.~P.~Schlenvoigt and R.~Sch\"utzhold,
``Heisenberg limit for detecting vacuum birefringence,''
Phys. Rev. D \textbf{101}, no.11, 116019 (2020)
[arXiv:2003.10519 [hep-ph]].

\bibitem{Schlenvoigt:2016jrd}
H.~P.~Schlenvoigt, T.~Heinzl, U.~Schramm, T.~E.~Cowan and R.~Sauerbrey,
``Detecting vacuum birefringence with x-ray free electron lasers and high-power optical lasers: a feasibility study,''
Phys. Scripta \textbf{91}, no.2, 023010 (2016)

\bibitem{Heinzl:2006xc}
T.~Heinzl, B.~Liesfeld, K.~U.~Amthor, H.~Schwoerer, R.~Sauerbrey and A.~Wipf,
``On the observation of vacuum birefringence,''
Opt. Commun. \textbf{267}, 318-321 (2006)
[arXiv:hep-ph/0601076 [hep-ph]].

\bibitem{Titov:2021kbj}
A.~I.~Titov, U.~H.~Acosta and B.~Kampfer,
``Positron energy distribution in a factorized trident process,''
Phys. Rev. A \textbf{104}, no.6, 062811 (2021)
[arXiv:2108.13043 [hep-ph]].

\bibitem{Tang:2022ixj}
S.~Tang and B.~King,
``Locally monochromatic two-step nonlinear trident process in a plane wave,''
[arXiv:2211.13299 [hep-ph]].

\bibitem{Ilderton:2010wr}
A.~Ilderton,
``Trident pair production in strong laser pulses,''
Phys. Rev. Lett. \textbf{106}, 020404 (2011)
[arXiv:1011.4072 [hep-ph]].

\bibitem{Boca:2012pz}
M.~Boca, V.~Dinu and V.~Florescu,
``Electron distributions in nonlinear Compton scattering,''
Phys. Rev. A \textbf{86}, 013414 (2012)
[arXiv:1206.6971 [physics.atom-ph]].

\bibitem{Roshchupkin:2022wpw}
S.~P.~Roshchupkin and S.~S.~Starodub,
``The effect of generation of narrow ultrarelativistic beams of positrons (electrons) in the process of resonant photoproduction of pairs on nuclei in a strong electromagnetic field,''
Laser Phys. Lett. \textbf{19}, no.11, 115301 (2022)
[arXiv:2205.14119 [hep-ph]].

\bibitem{Roshchupkin:2022hfq}
S.~Roshchupkin, A.~Dubov and S.~Starodub,
``Radiation of High-Energy Gamma Quanta by Ultrarelativistic Electrons on Nuclei in Strong X-ray Fields,''
Universe \textbf{8}, no.4, 218 (2022)
[arXiv:2202.08151 [hep-ph]].

\bibitem{Roshchupkin:2021vut}
S.~P.~Roshchupkin, A.~V.~Dubov, V.~V.~Dubov and S.~S.~Starodub,
``Fundamental physical features of resonant spontaneous bremsstrahlung radiation of ultrarelativistic electrons on nuclei in strong laser fields,''
New J. Phys. \textbf{24}, no.1, 013020 (2022)
[arXiv:2108.10186 [hep-ph]].

\bibitem{Dubov:2020kwi}
A.~Dubov, V.~V.~Dubov and S.~P.~Roshchupkin,
Universe \textbf{6}, no.9, 143 (2020)
doi:10.3390/universe6090143

\bibitem{Bos:1978th}
J.~Bos, W.~Brock, H.~Mitter and T.~Schott,
``RESONANCES AND INTENSITY DEPENDENT SHIFTS OF THE MOLLER CROSS-SECTION IN A STRONG LASER FIELD,''
J. Phys. A \textbf{12}, 715 (1979)

\bibitem{Sizykh:2021ywt}
G.~K.~Sizykh, S.~P.~Roshchupkin and V.~V.~Dubov,
``Resonant Effect of High-Energy Electron\textendash{}Positron Pairs Production in Collision of Ultrarelativistic Electrons with an X-ray Electromagnetic Wave,''
Universe \textbf{7}, no.7, 210 (2021)

\end{thebibliography}
\end{document}